\documentclass[prc,superscriptaddress,twocolumn,showpacs]{revtex4-1}
\usepackage{epsfig}
\usepackage{psfrag}
\usepackage{amsmath}
\usepackage{sistyle}

\begin{document}

\title{Study of bound states in $^{12}$Be through low-energy $^{11}$Be(d,p)-transfer reactions}

\author{J. G. Johansen}
\affiliation{Department of Physics and Astronomy, Aarhus University, 
DK-8000 Aarhus C, Denmark}
\affiliation{Institut f\"ur Kernphysik, Technische Universit\"at Darmstadt, D-64289 Darmstadt, Germany}
\author{ V. Bildstein}
\affiliation{Department of Physics, University of Guelph, Guelph, Ontario, Canada N1G2W1}
\author{M. J. G. Borge}
\affiliation{CERN, CH-1211 Gen\`eve, Switzerland}
\affiliation{Instituto de Estructura de la Materia, CSIC, E-28006 Madrid, Spain} 
\author{M. Cubero}
\affiliation{Instituto de Estructura de la Materia, CSIC, E-28006 Madrid, Spain} 
\author{J. Diriken}
\affiliation{Instituut voor Kern- en Stralingsfysica, KU Leuven, 3001 Leuven, Belgium}
\affiliation{Belgian Nuclear Research Centre SCK$\cdot$CEN, Boeretang 200, B-2400 Mol, Belgium}
\author{J. Elseviers}
\affiliation{Instituut voor Kern- en Stralingsfysica, KU Leuven, 3001 Leuven, Belgium}
\author{L. M. Fraile}
\affiliation{Grupo de Fisica  Nuclear, Universidad Complutense, CEI Moncloa, E-28040 Madrid, Spain}
\author{H. O. U. Fynbo}
\affiliation{Department of Physics and Astronomy, Aarhus University, 
DK-8000 Aarhus C, Denmark}
\author{L. P. Gaffney}
\affiliation{Oliver Lodge Laboratory, University of Liverpool, L69 7ZE, England, UK}
\author{R. Gernh\"auser}
\affiliation{Physik Department E12, Technische Universit\"at M\"unchen, 85748 Garching, Germany}
\author{B. Jonson}
\affiliation{Fundamental Fysik, Chalmers Tekniska H\"ogskola, S-41296 G\"oteborg, Sweden}
\author{G. T. Koldste}
\affiliation{Department of Physics and Astronomy, Aarhus University, 
DK-8000 Aarhus C, Denmark}
\author{J. Konki}
\altaffiliation[Present address: ]{University of Jyv\"askyl\"a, Jyv\"askyl\"a FI-40014, Finland}
\affiliation{CERN, CH-1211 Gen\`eve, Switzerland}
\affiliation{Helsinki Institute of Physics, PO box 64, FI-00014 Helsinki, Finland}
\author{T. Kr\"oll}
\affiliation{Institut f\"ur Kernphysik, Technische Universit\"at Darmstadt, D-64289 Darmstadt, Germany}
\author{R. Kr\"ucken}
\affiliation{Physik Department E12, Technische Universit\"at M\"unchen, 85748 Garching, Germany}
\affiliation{TRIUMF, Vancouver BC, V6T 2A3, Canada}
\author{D. M\"ucher}
\affiliation{Physik Department E12, Technische Universit\"at M\"unchen, 85748 Garching, Germany}
\author{T. Nilsson}
\affiliation{Fundamental Fysik, Chalmers Tekniska H\"ogskola, S-41296 G\"oteborg, Sweden}
\author{K. Nowak}
\affiliation{Physik Department E12, Technische Universit\"at M\"unchen, 85748 Garching, Germany}
\author{J. Pakarinen}
\altaffiliation[Present address: ]{University of Jyv\"askyl\"a, Jyv\"askyl\"a FI-40014, Finland}
\affiliation{CERN, CH-1211 Gen\`eve, Switzerland}
\affiliation{Helsinki Institute of Physics, PO box 64, FI-00014 Helsinki, Finland}
\author{V. Pesudo}
\affiliation{Instituto de Estructura de la Materia, CSIC, E-28006 Madrid, Spain} 
\author{R. Raabe}
\affiliation{Instituut voor Kern- en Stralingsfysica, KU Leuven, 3001 Leuven, Belgium}
\author{K. Riisager}
\affiliation{Department of Physics and Astronomy, Aarhus University, 
DK-8000 Aarhus C, Denmark}
\author{M. Seidlitz}
\affiliation{Institut f\"ur Kernphysik, Universit\"at zu K\"oln, D-50937 K\"oln, Germany}
\author{O. Tengblad}
\affiliation{Instituto de Estructura de la Materia, CSIC, E-28006 Madrid, Spain} 
\author{H. T\"ornqvist}
\affiliation{CERN, CH-1211 Gen\`eve, Switzerland}
\affiliation{Fundamental Fysik, Chalmers Tekniska H\"ogskola, S-41296 G\"oteborg, Sweden}
\author{D. Voulot}
\affiliation{CERN, CH-1211 Gen\`eve, Switzerland}
\author{N. Warr}
\affiliation{Institut f\"ur Kernphysik, Universit\"at zu K\"oln, D-50937 K\"oln, Germany}
\author{F. Wenander}
\affiliation{CERN, CH-1211 Gen\`eve, Switzerland}
\author{K. Wimmer}
\affiliation{Physik Department E12, Technische Universit\"at M\"unchen, 85748 Garching, Germany}
\affiliation{Department of Physics, Central Michigan University, Mount Pleasant, Michigan 48859, USA}
\author{H. De Witte}
\affiliation{Instituut voor Kern- en Stralingsfysica, KU Leuven, 3001 Leuven, Belgium}

\date{\today}

\begin{abstract}
  The bound states of $^{12}$Be have been studied through a
  $^{11}$Be(d,p)$^{12}$Be transfer reaction experiment in inverse
  kinematics.  A \SI{2.8}{MeV/u} beam of $^{11}$Be was produced using the
  REX-ISOLDE facility at CERN.  The outgoing protons were detected with
  the T-REX silicon detector array. The MINIBALL germanium array was used to detect gamma rays from the excited states in $^{12}$Be. The gamma-ray detection
  enabled a clear identification of the four known bound states in
  $^{12}$Be, and each of the states has been studied individually. Differential
  cross sections over a large angular range have been
  extracted. Spectroscopic factors for each of the states have been
  determined from DWBA calculations and have been compared to previous
  experimental and theoretical results.
\end{abstract}


\maketitle

\section{Introduction}
The structure of the light neutron-rich nuclei has presented many
challenges during the last decades \cite{Jon04} and this
area of the nuclear chart is a prime region
for investigations of halos \cite{Jen04,Tani13,Riis13}, cluster states
\cite{Oer06}, unbound systems \cite{Baum12,Simon13} as well as the
vanishing of shells \cite{Kanun13}. A key question in these
topics is the spectroscopic composition of the bound states, which can
be accessed experimentally in complementary ways
\cite{Blaum13,Auman13,Kee07}. We are here concerned with the structure
of the bound states in $^{12}$Be. The states are probed via transfer reactions.  This method has recently been employed also for the study of other exotic nuclei
\cite{Wimmer10,Bonac13,Jones13}. In neither of the neighboring isotopes
$^{11}$Li and $^{11}$Be can the ground states be written as a simple
single-particle configuration (see \cite{Schm12} and the references
above). This seems to also be the case for $^{12}$Be.

\begin{figure}[h]
	\includegraphics[width=.45\textwidth]{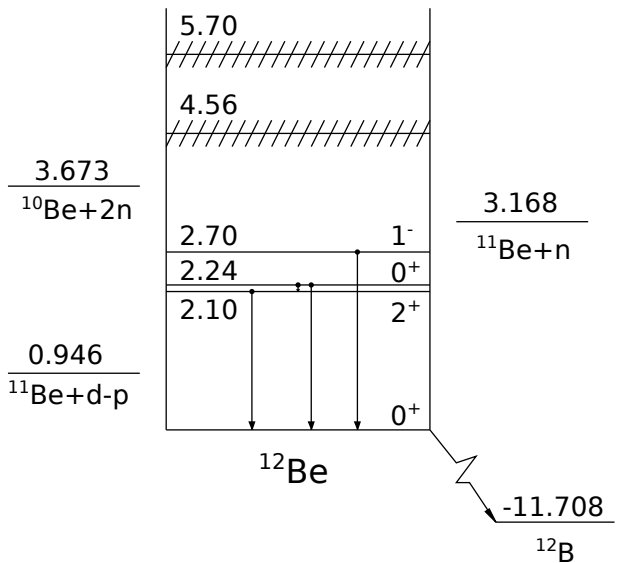}
	\caption{The level scheme of $^{12}$Be containing the known states and 
		 resonances as well as the gamma decay transitions for the 
		 bound states. The energies are given in MeV and the values for
		 the bound states and the resonances are taken from 
                 \cite{Iwasaki00,Shimoura03,Shimoura07,Fort94}.}
	\label{fig:level}
\end{figure}

Four bound states are presently known in $^{12}$Be, see Fig.~\ref{fig:level}. The highest lying
bound state $(1^-_1)$ has an excitation energy of only $E^*=\SI{2.70}{MeV}$. Hence the level density in $^{12}$Be is relatively high for a light nucleus, in comparison the first excited states
in $^{10}$Be and $^{13}$B have an excitation energy above $\SI{3}{MeV}$ and in $^{12}$C the first 
excited state is at $\SI{4.4}{MeV}$. This high level density is believed to be due to 
configurations from both the $0p_{1/2}$-shell and the $1s_{1/2}0d_{5/2}$-shell, a situation which is also known to occur in 11Be \cite{Tal60}. The configuration mixing was first suggested in 1976 by Barker \cite{Bark76} following a $\beta$-decay study
of $^{12}$Be. Further experimental support for this suggestion has come
through measurements of transition strengths to the $1^-_1$
\cite{Iwasaki00}, $0^+_2$ \cite{Shimoura03,Shimoura07} and $2^+_1$
bound states \cite{Imai09}, through extraction of the ground state
charge radius \cite{Kri12} and through nuclear knock out \cite{Navin00}, 
break-up \cite{Pain06,Peters11}, transfer \cite{Kanungo10} and charge
exchange \cite{Mehar12} reactions.  These measurements have shown that
the N=8 magic number is clearly
broken in $^{12}$Be and the detailed mixing of the shells is still
being investigated. The short lifetime combined with the narrow
separation of the bound states has made it difficult to study these
states individually.

$^{11}$Be is a well known 1-neutron halo nucleus. This large separation of the $^{10}$Be and the halo-neutron in $^{11}$Be has led to an
interpretation of $^{12}$Be as having a three-particle structure with a
$^{10}$Be core and two neutrons, e.g.\ \cite{Rome08}. The $1^-_1$ state is particularly interesting in a three-body model. This state is only $\SI{1}{MeV}$ below the two-neutron threshold and is expected to be a two-neutron halo in the three-body model with one
neutron in an $s$ state and the other in a $p_{1/2}$ state. A fifth
bound state that differs from the $1^-_1$ state only in spin coupling has been suggested: a
$0^-_1$ state with an excitation energy around $E^*= \SI{2.5}{MeV}$
\cite{Rome208}. However, this state has never been seen
experimentally. 

Several other models have been used to
predict the $^{12}$Be structure: apart from shell models and a simple
wavefunction ansatz \cite{Fortune12}, antisymmetrized molecular
dynamics \cite{Kan03}, the deformed potential model \cite{Ham07}, 
as well as the generator coordinate method and no core shell 
model \cite{Duf10} have all been employed.

The first studies of excited states in $^{12}$Be were in transfer
reactions, mainly $^{10}$Be(t,p)$^{12}$Be, these results are
summarized in Fortune et al. \cite{Fort94}. Most recent studies have
been in break-up reactions, and also a $^{11}$Be(d,p)$^{12}$Be transfer
reaction has been performed at TRIUMF by R. Kanungo et
al. \cite{Kanungo10}. Spectroscopic factors were determined in the
latter for all four bound states. The value for the $0^+_2$ state was
only given with a large uncertainty, due to the inability to clearly
distinguish it from the $2^+_1$ state. The
spectroscopic factors determined in the experiment at TRIUMF have
later been questioned, since it disagreed with theoretic calculations
 \cite{Fortune12}.

In this paper, we report on a $^{11}$Be(d,p)-experiment performed
at ISOLDE. The set-up represents an improvement upon an earlier
$^9$Li(d,p) experiment \cite{Jep05} as both gamma rays and charged
particles were measured, enabling a clear
identification of all the bound states in $^{12}$Be. Hence, detailed
studies of each state have been made and spectroscopic factors have
been determined for all the four states. The lifetime and the
branching ratio of the decay of the $0^+_2$ state have also been
determined. Results from the other reaction channels as well as on the
unbound resonances in $^{12}$Be \cite{Gar12} will be reported elsewhere.

The paper starts with a description of the experimental setup and
experimental procedure, section~\ref{S:Experiment}. The analysis of
the data is described in section~\ref{S:Analysis}. The focus of the
analysis is the identification of the individual states, but the
lifetime of and the branching ratio for the decay of the $0^+_2$ state
is also given. The analysis is done in three steps described in
section~\ref{S:Analysis}. The determined differential cross sections
are presented along with the DWBA calculations in
section~\ref{S:Results}. The spectroscopic factors are also presented
and discussed in this section. The paper ends with a short summary and
conclusion in section~\ref{S:Conclusion}.

\section{The experimental procedure}
\label{S:Experiment}
The experiment was performed at the ISOLDE facility, CERN,
Switzerland. The $^{11}$Be activity was produced by a $\SI{1.4}{GeV}$ proton
beam through fragmentation of a uranium carbide target. The Be atoms
were subsequently ionized via laser ionization \cite{Fed00}, mass
separated and led to the REX-ISOLDE post-accelerator. Here they were
bunched in REXTRAP, fully stripped to charge state +4 in REXEBIS and
finally post-accelerated to $\SI{2.8}{MeV/u}$ ($\SI{30.7}{MeV}$) 
in the REX linear accelerator \cite{Kester03}. The beam intensity after 
post-acceleration fluctuated between $4.4 \times 10^6$/s and $1 
\times 10^7$/s. This led to a total number of $^{11}$Be nuclei of
$N_{^{11}\text{Be}} = 1.11(25) \times 10^{12}$. The beam intensity was
determined by Coulomb scattering on a silver target, which was performed regularly during the experiment. The fluctuation
in the beam intensity was monitored via the rate of detected particles
(p, d and t) throughout the experiment. The set-up allowed for a study of several properties of the secondary $^{11}$Be beam. The beam spot was determined to be a flat distribution on an area with a diameter of approximately $\SI{6}{mm}$, details are given in \cite{Johansen13}.

A deuterated polyethylene (CD$_2$) target was used in the
experiment. The thickness of the target was
1.00(5) mg/cm$^2$. Runs on a pure carbon target and a regular
polyethylene target (CH$_2$) were performed and provided information
about reactions on C and H in the primary target.
\begin{figure*}[t]
	\begin{center}
	\includegraphics[scale=0.8]{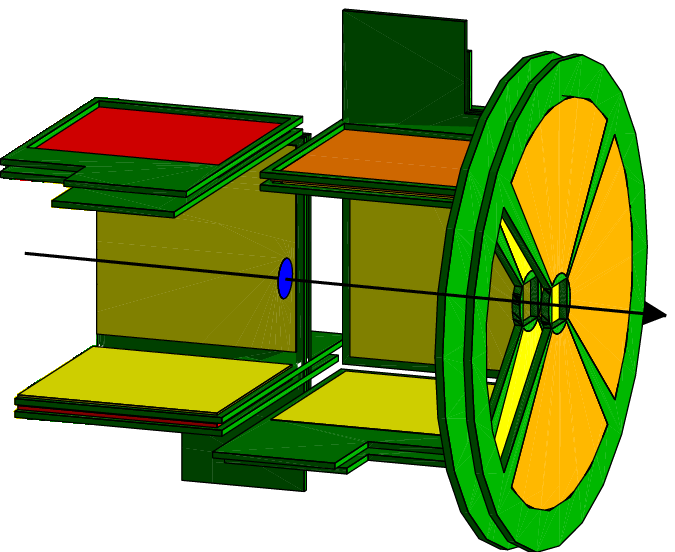}
	\hspace{1.5cm}
	\includegraphics[width=.5\textwidth,height=5cm]{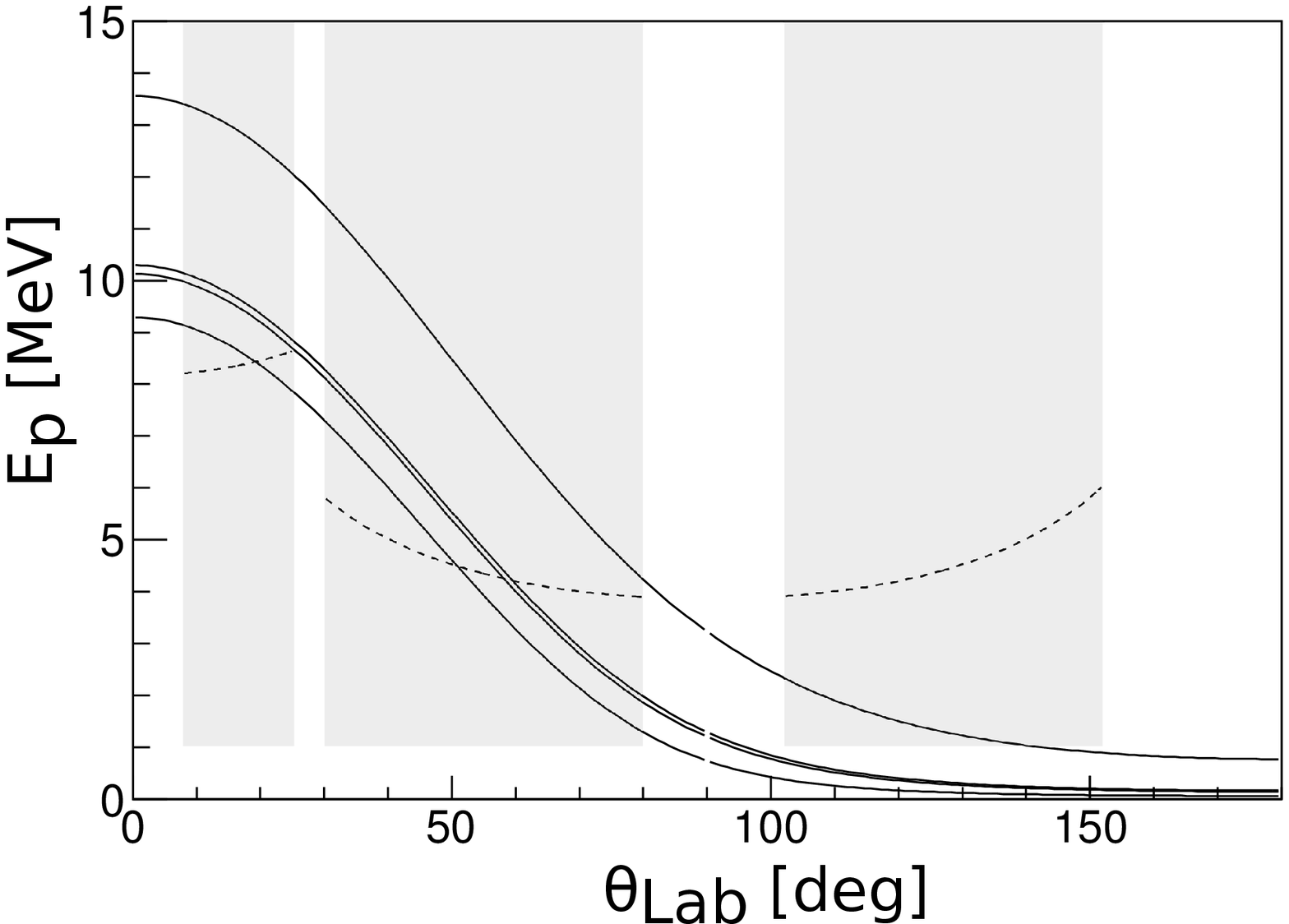}\\
	\vspace{-4.5cm}
	\hspace{3.8cm} (a) \hspace{.5\textwidth} (b)
	\\
	\vspace{4.5cm}
	\end{center}
	\caption{(Color online) (a): A drawing of the T-REX silicon array \cite{Bildstein12}. The three
          detectors on the left side are omitted to clear the view
          inside the T-REX. (b): The kinetic curves for the emitted
          protons in a $^{11}$Be(d,p)$^{12}$Be reaction (solid
          lines). Each line corresponds to a population of one of
	  the four bound states, Fig.~\ref{fig:level}. The gray area represents 
	  the angles and energies
          covered by the T-REX and the dashed line corresponds to the
          minimum energy required for $\Delta E-E$-identification.}
	\label{fig:TREX}
\end{figure*}

A setup specialized for transfer reaction experiments at ISOLDE was
used. The setup consisted of the MINIBALL germanium detector array
\cite{Eberth01,Warr13} in combination with the T-REX silicon detector setup
\cite{Bildstein12}. The T-REX was used to detect the light charged
particles from the reaction. The T-REX consisted
of 12 silicon telescope detectors placed to cover
angles from $8^\text{o}$ to $152^\text{o}$ in the laboratory and
with an almost $2\pi$-azimuthal angular coverage. A drawing of the
T-REX is seen in Fig.~\ref{fig:TREX}a. Fig.~\ref{fig:TREX}b shows the
angles and energies covered by the T-REX (grey area). The dashed lines
represents the energy required for a proton to pass through the first of the telescope
detectors. The kinetic curves of the four known bound states are also
shown in Fig.~\ref{fig:TREX}b. Particle identification through $\Delta
E-E$ plots can be performed above the dashed lines. Particles with
energy less than $\SI{1}{MeV}$ could not be separated from the noise level.

The gamma-ray detection provided by the MINIBALL was required to separate the bound states in $^{12}$Be. The MINIBALL
consists of 24 germanium detectors placed in eight clusters. The
clusters were placed to cover a wide angular range. The germanium detectors
had an energy range up to $\SI{8}{MeV}$.
The energy-dependent detection
efficiency was determined using three gamma sources ($^{152}$Eu,
$^{60}$Co and $^{207}$Bi) and gamma rays from $\beta$-decay
of $^{11}$Be. The $^{11}$Be beam used for the efficiency calculation
was stopped in an aluminum foil at the target position. The detection 
efficiencies for decays occurring at the target position are given,
 for the relevant decays, in Table~\ref{tab:Efficiency}. More
details on the experimental procedures can be found in
\cite{Johansen12}. 
\begin{table}[b]
\center
\begin{tabular}{|c|c|c|}
\hline
Decay & $E_\gamma$ [keV] & $\epsilon$  [$\%$] \\
\hline
$0^+_2 \rightarrow 2^+_1$ & 144 & 16.2(5) \\
$0^+_2 \rightarrow 0^+_1$ & 511 (pair creation) & 8.2(5)\\
$2^+_1 \rightarrow 0^+_1$ & 2107 & 3.5(2)\\
$1^-_1 \rightarrow 0^+_1$ & 2680 & 3.0(2) \\
\hline
\end{tabular}
\caption{The $\gamma$-energy and the MINIBALL detection efficiency for 
the four main $\gamma$-decay lines in $^{12}$Be.}
\label{tab:Efficiency}
\end{table}
\begin{figure}[t]
	\psfrag{dE}[cc]{{\normalsize $\Delta E$ [MeV]}}
	\psfrag{E}[bb]{{\normalsize $E$ [MeV]}}
	\includegraphics[width=0.5\textwidth]{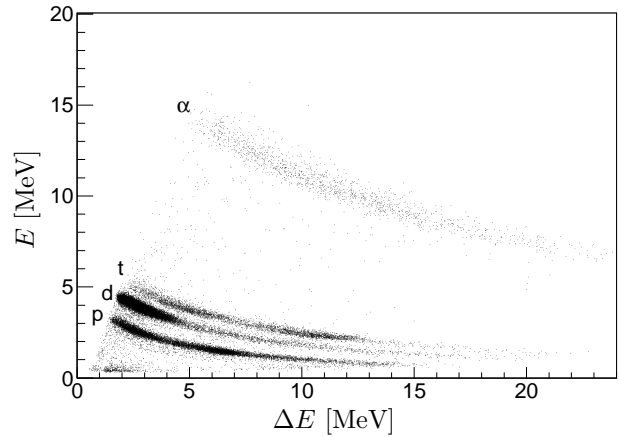}
	\caption{A $\Delta E-E$-plot for a strip in one of the detectors
          covering the forward laboratory angles. The three curves 
	  corresponding to p, d and t are clearly separated.}
	\label{fig:dEE}
\end{figure}

\begin{figure*}[t]
	\begin{center}
	\psfrag{N}[bb]{{\normalsize counts/$\SI{50}{keV}$}}
	\psfrag{E}[cc]{{\normalsize $E^*$ [MeV]}}
	\includegraphics[width=0.45\textwidth]{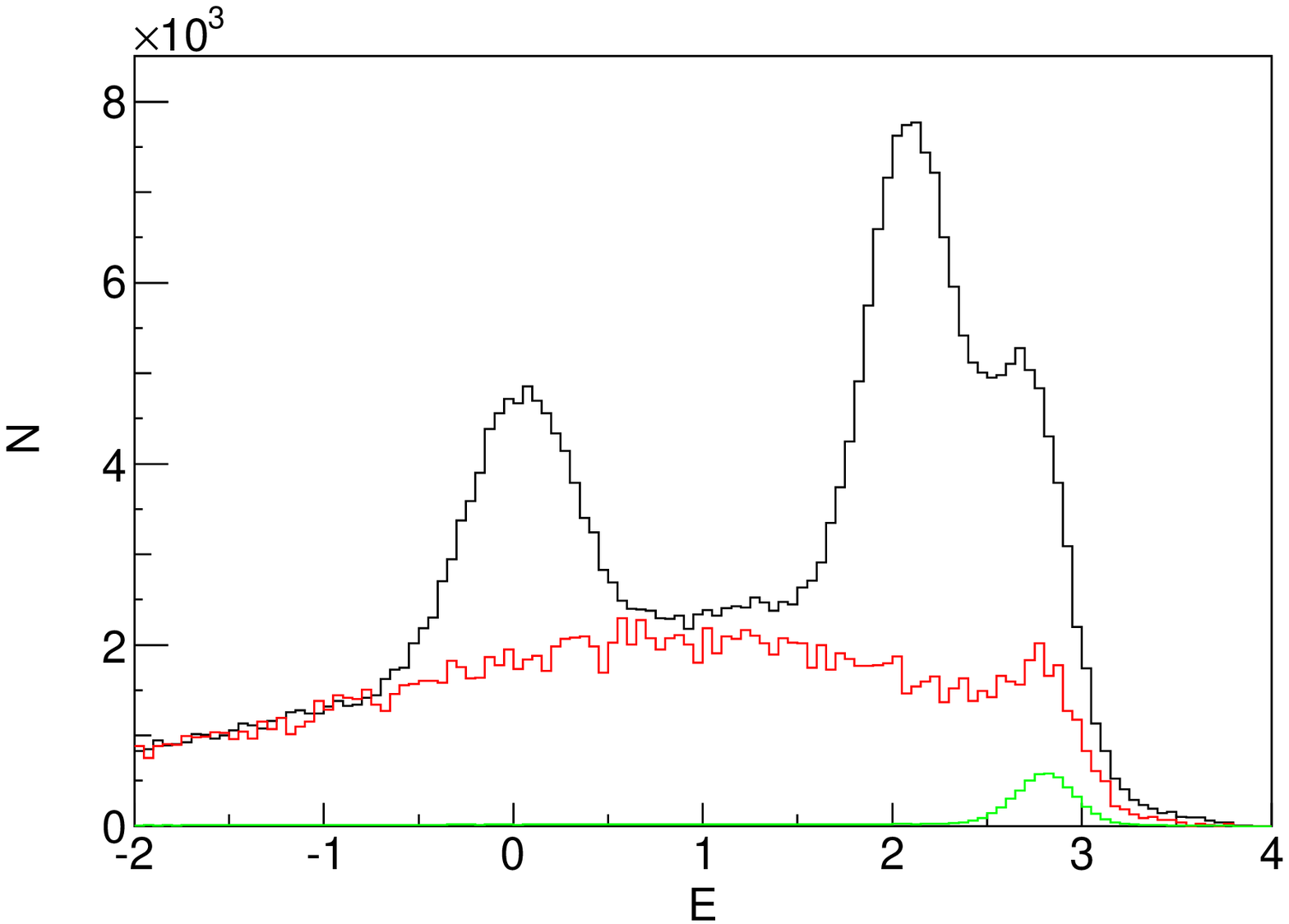}
	\includegraphics[width=0.45\textwidth]{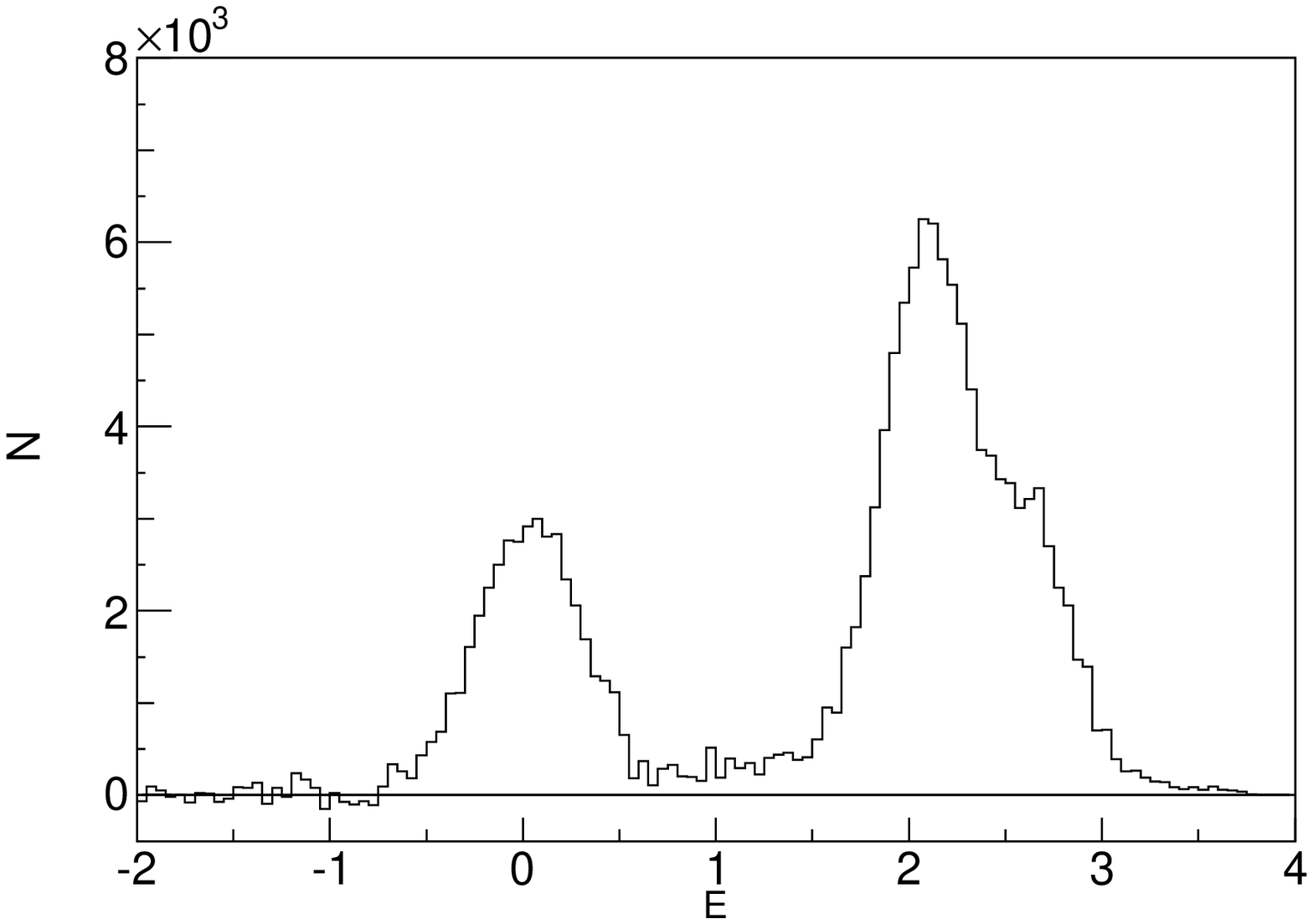}\\
	\vspace{-4.8cm}
	\hspace{5.7cm} (a) \hspace{.42\textwidth} (b)
	\\
	\vspace{4.8cm}
	\includegraphics[width=0.45\textwidth]{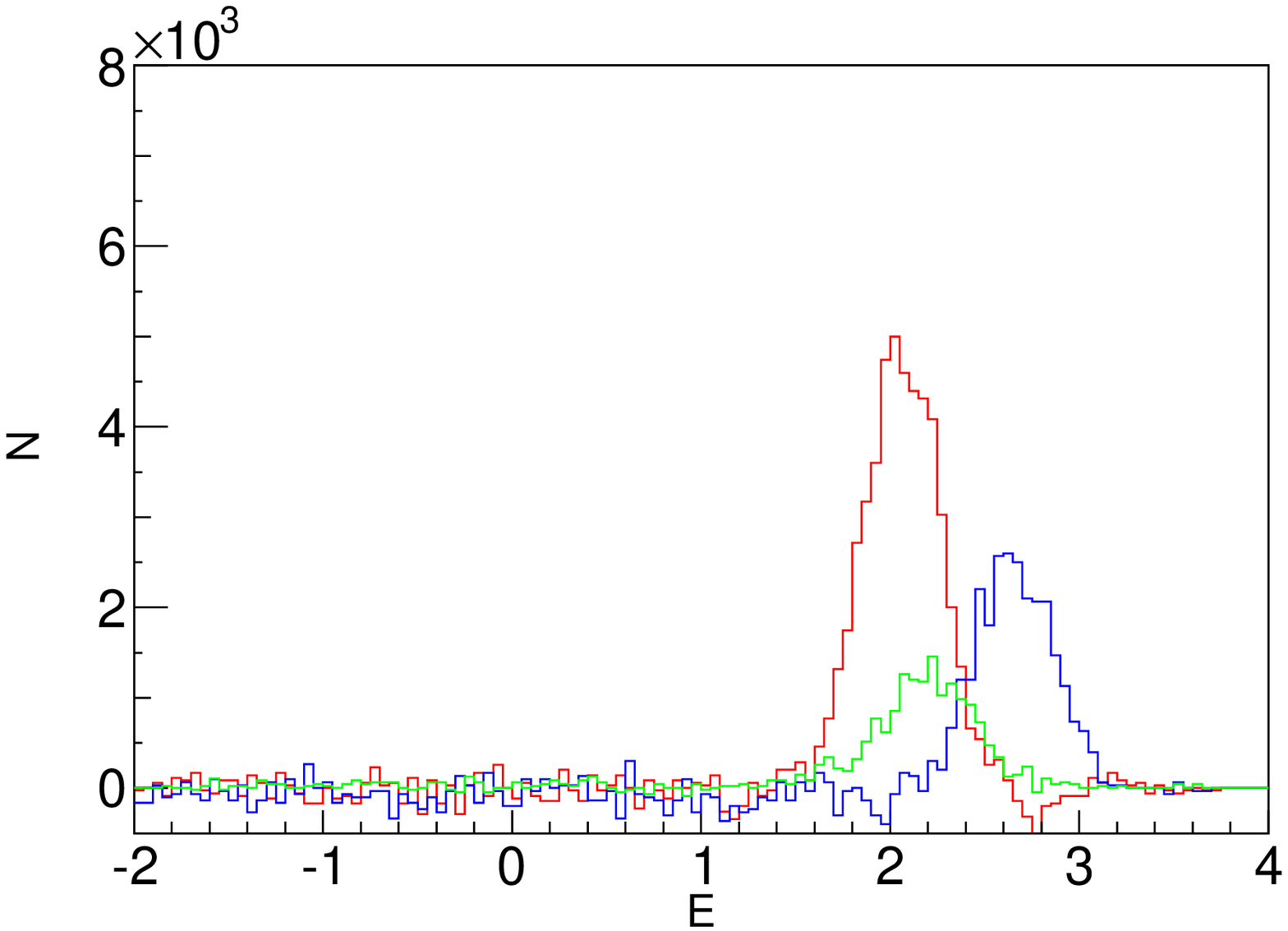}
	\includegraphics[width=0.45\textwidth]{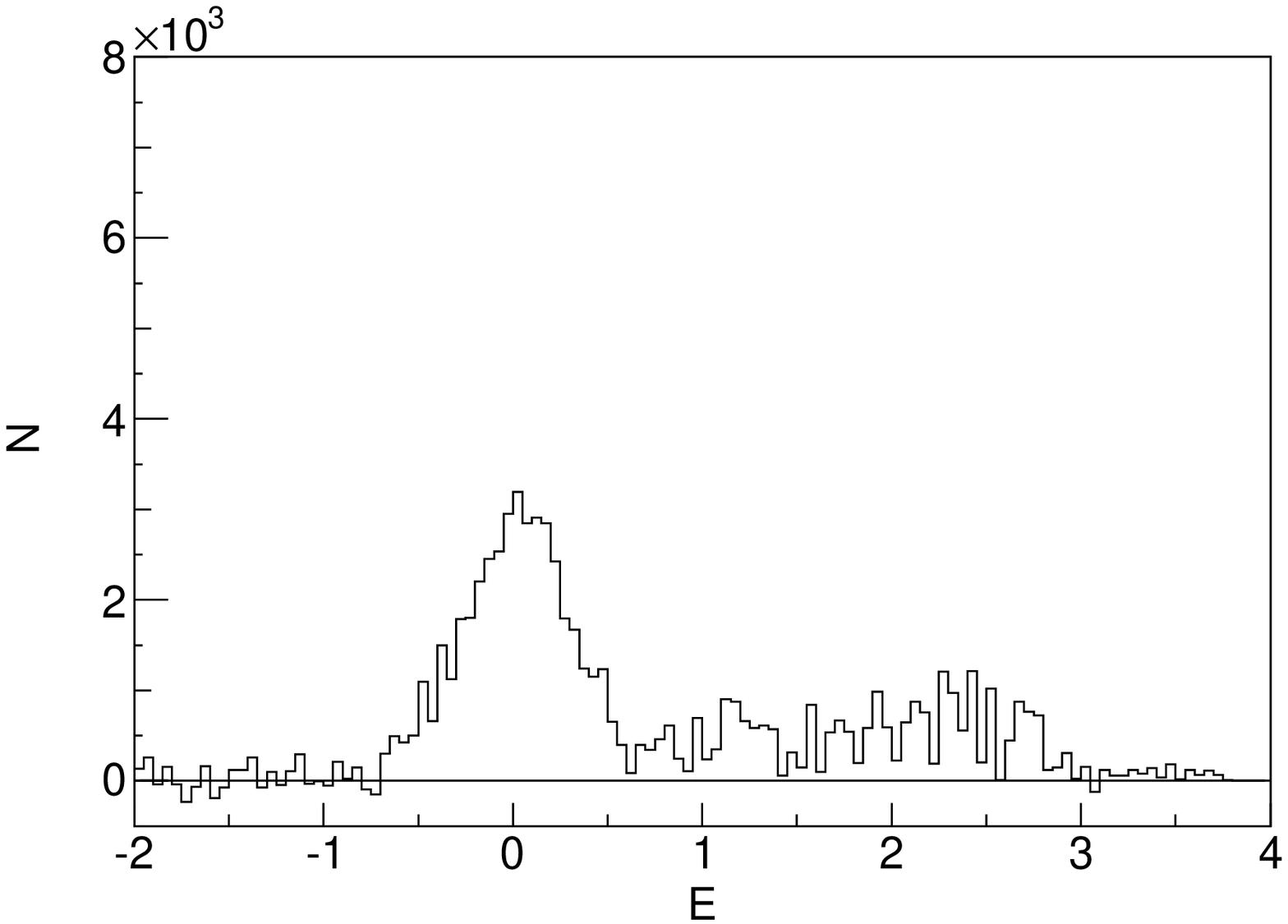}\\
	\vspace{-4.8cm}
	\hspace{5.7cm} (c) \hspace{.42\textwidth} (d)
	\\
	\vspace{4.8cm}
	\end{center}
	\caption{(Color online) Four plots showing the excitation energy of $^{12}$Be.
		 (a): The excitation energy spectrum determined from the
		 momentum of protons identified from the $\Delta E-E$-plots 
		 (Black) along with the background spectra determined 
		 from runs on the carbon target (red (dark gray)) and the CH$_2$ 
		 target (green (light gray)). (b): The excitation energy spectrum from (a)
		 with the two background spectra subtracted. (c): Excitation
		 energy spectra determined from protons gated on the gamma-lines
		 shown in Fig.~\ref{fig:EgammaHE}; red (dark gray): $2^+_1$ 
		 ($E_\text{peak}=\SI{2061}{keV}$), green (light gray): $0^+_2$ 
		 ($E_\text{peak}=\SI{2190}{keV}$) and blue (black): $1^-_1$
		 ($E_\text{peak}=\SI{2658}{keV}$). (d): The excitation energy 
		 spectrum from (b) with the three spectra gated on gamma rays
		 in (c) subtracted only the ground state peak is present.
		}
	\label{fig:Eex}
\end{figure*}
\section{Identification of the bound states in $^{12}$Be}
\label{S:Analysis}
The identification of the four bound states in $^{12}$Be is
performed in three steps. The protons are identified in a $\Delta E-E$
plot if the energy of the particles is sufficiently high
(section~\ref{SS:HE}). This is only the case for the forward
laboratory angles according to Fig.~\ref{fig:TREX}b. The excited
states are identified using gating on gamma-ray energies 
 afterwards. The identification of (d,p)-reactions from particles stopped
in the $\Delta E$-detectors is divided in two, forward and backward angles.
In forward angles only protons populating $^{12}$Be in an excited state
will be stopped in the $\Delta E$-detector, Fig.~\ref{fig:TREX}b. 
These protons can be identified by gating on gamma-ray energies (section~\ref{SS:LEfor}).
Protons populating the ground state will have sufficient energy to
penetrate the $\Delta E$-detector in the forward angle and can be
ignored when analyzing particles stopped in the forward $\Delta E$-detectors.
In backward laboratory angles only protons populating the ground state 
and particles from reactions on carbon in the target have sufficient 
energy to be separated from the noise level. The latter can be taken into
account via the runs on a pure carbon target (section~\ref{SS:LEback}).
\begin{figure*}[t]
	\begin{center}
	\psfrag{N}[bb]{{\normalsize counts/keV}}
	\psfrag{E}[cc]{{\normalsize $E_\gamma$ [MeV]}}
	\psfrag{1+}[bc]{{\normalsize \hspace{1cm}$1^- \rightarrow 0^+_1$}}
	\psfrag{2+}[bc]{{\normalsize \hspace{1cm}$2^+ \rightarrow0^+_1$}}
	\includegraphics[width=0.45\textwidth]{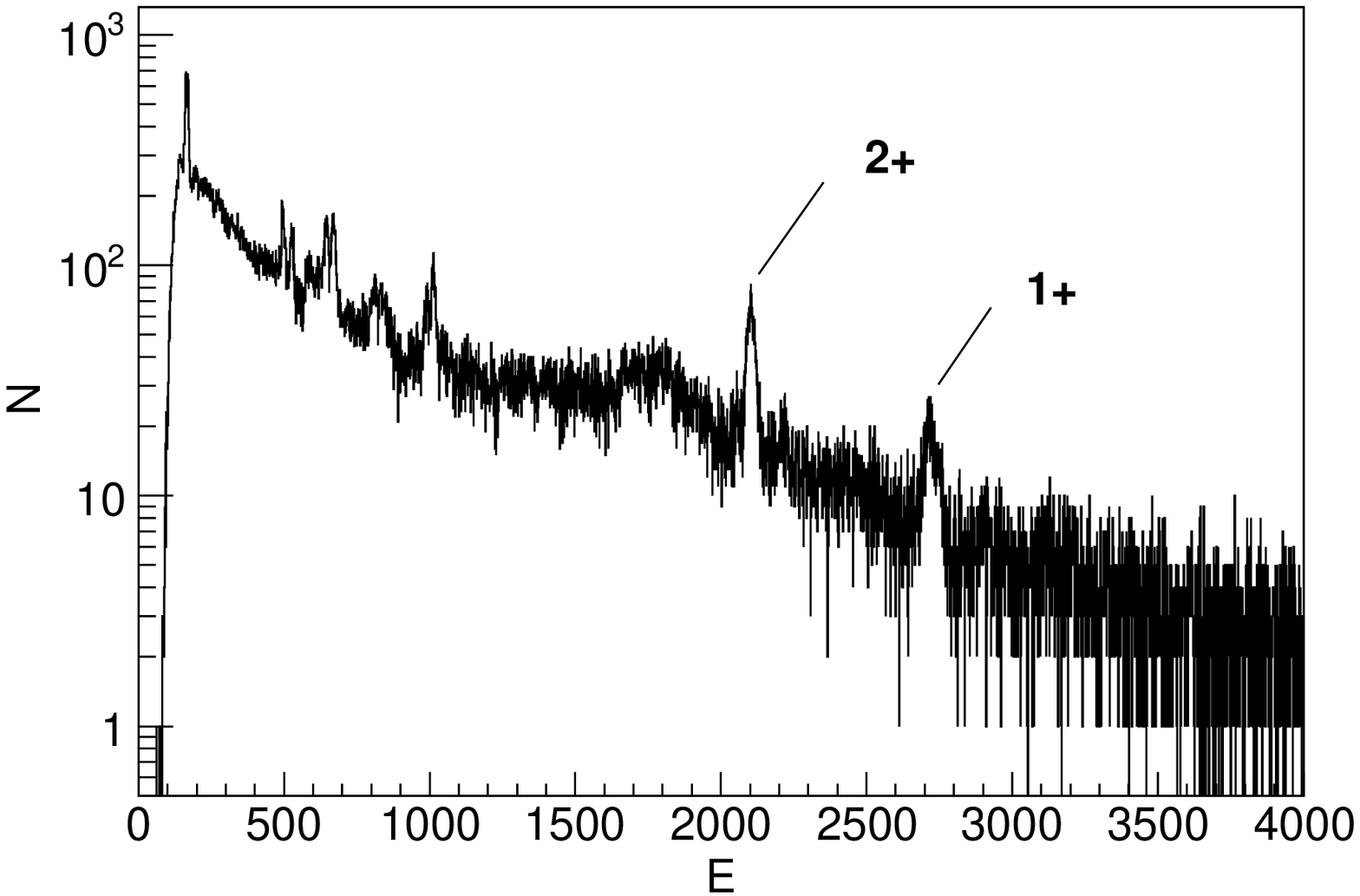}
	\psfrag{E1}[bb]{{\normalsize $E_\gamma$ [MeV]}}
	\psfrag{dt}[cc]{{\normalsize $\Delta t$ [ns]}}
	\includegraphics[width=0.45\textwidth]{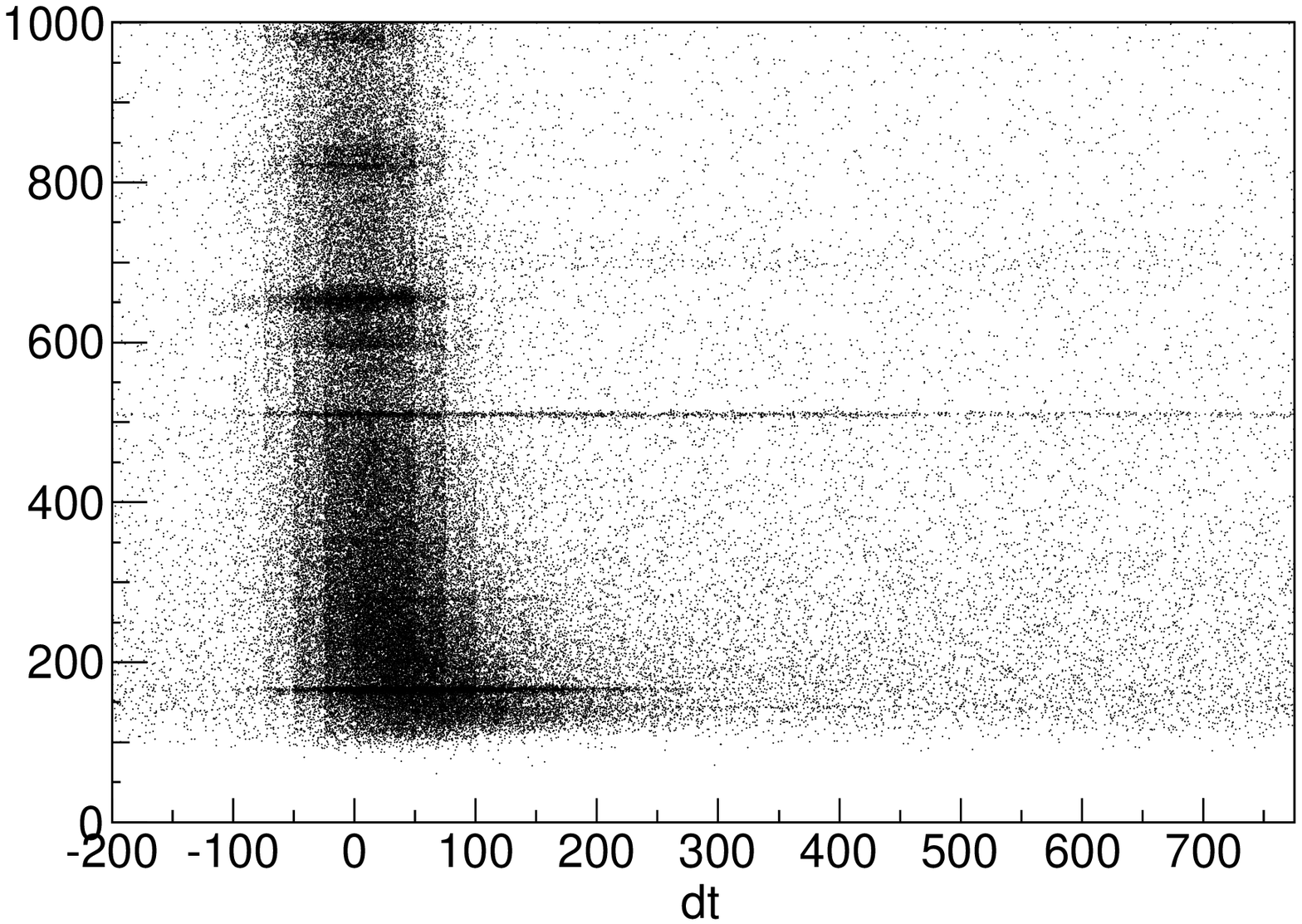}\\
	\vspace{-4.8cm}
	\hspace{5.7cm} (a) \hspace{.42\textwidth} (b)
	\\
	\vspace{4.8cm}
	\psfrag{N}[bb]{{\normalsize counts/keV}}
	\psfrag{E}[cc]{{\normalsize $E_\gamma$ [MeV]}}
	\psfrag{0+1}[bc]{{\normalsize \hspace{1cm}$0^+_2 \rightarrow 2^+$}}
	\psfrag{0+2}[tb]{{\normalsize \hspace{1cm}$0^+_2 \rightarrow0^+_1$}}
	\psfrag{Ge}[bc]{{\normalsize $^{72}$Ge$^*$}}
	\includegraphics[width=0.45\textwidth]{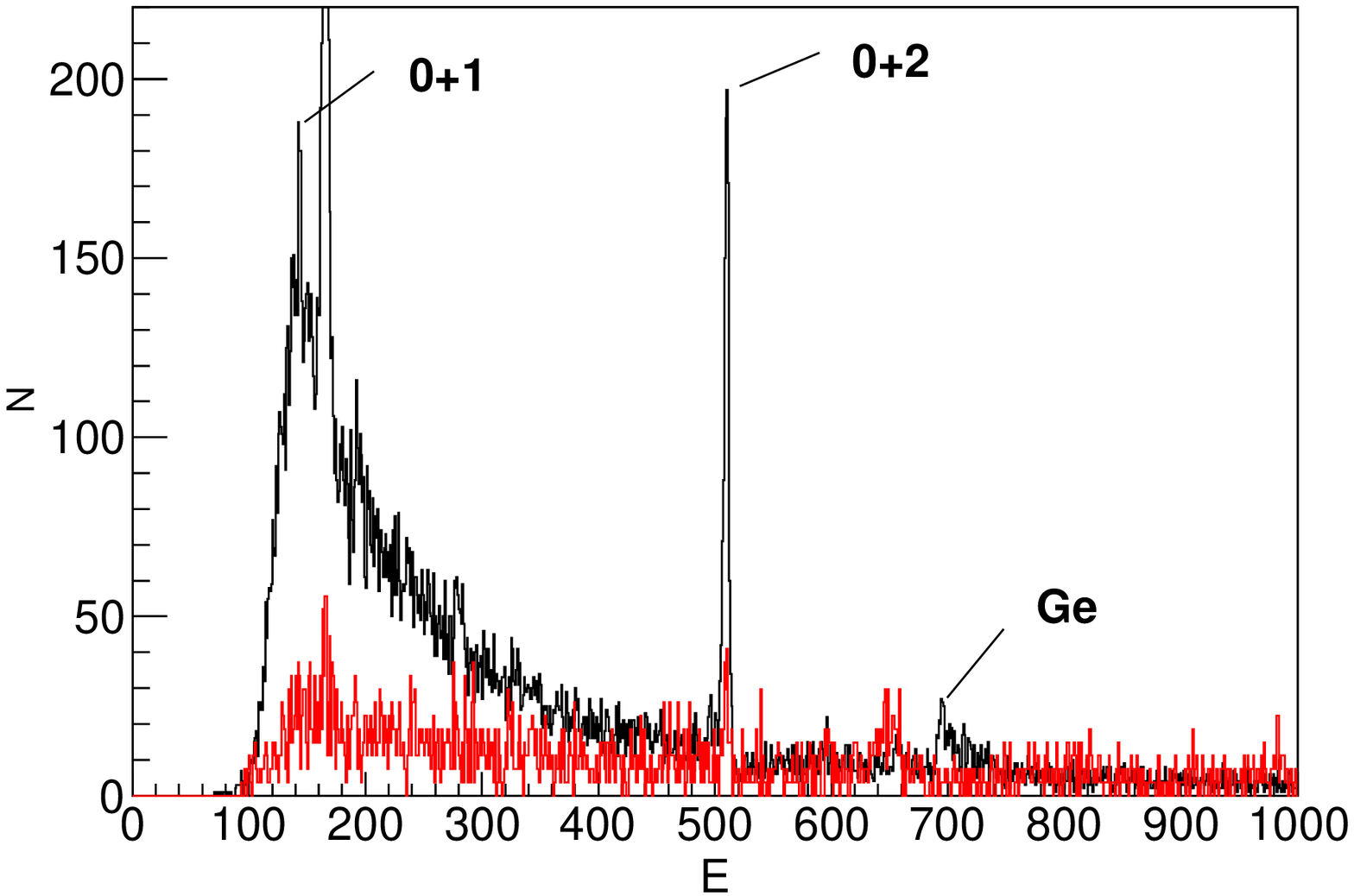}
	\psfrag{N}[bb]{{\normalsize counts/$\SI{25}{ns}$}}
	\includegraphics[width=0.45\textwidth]{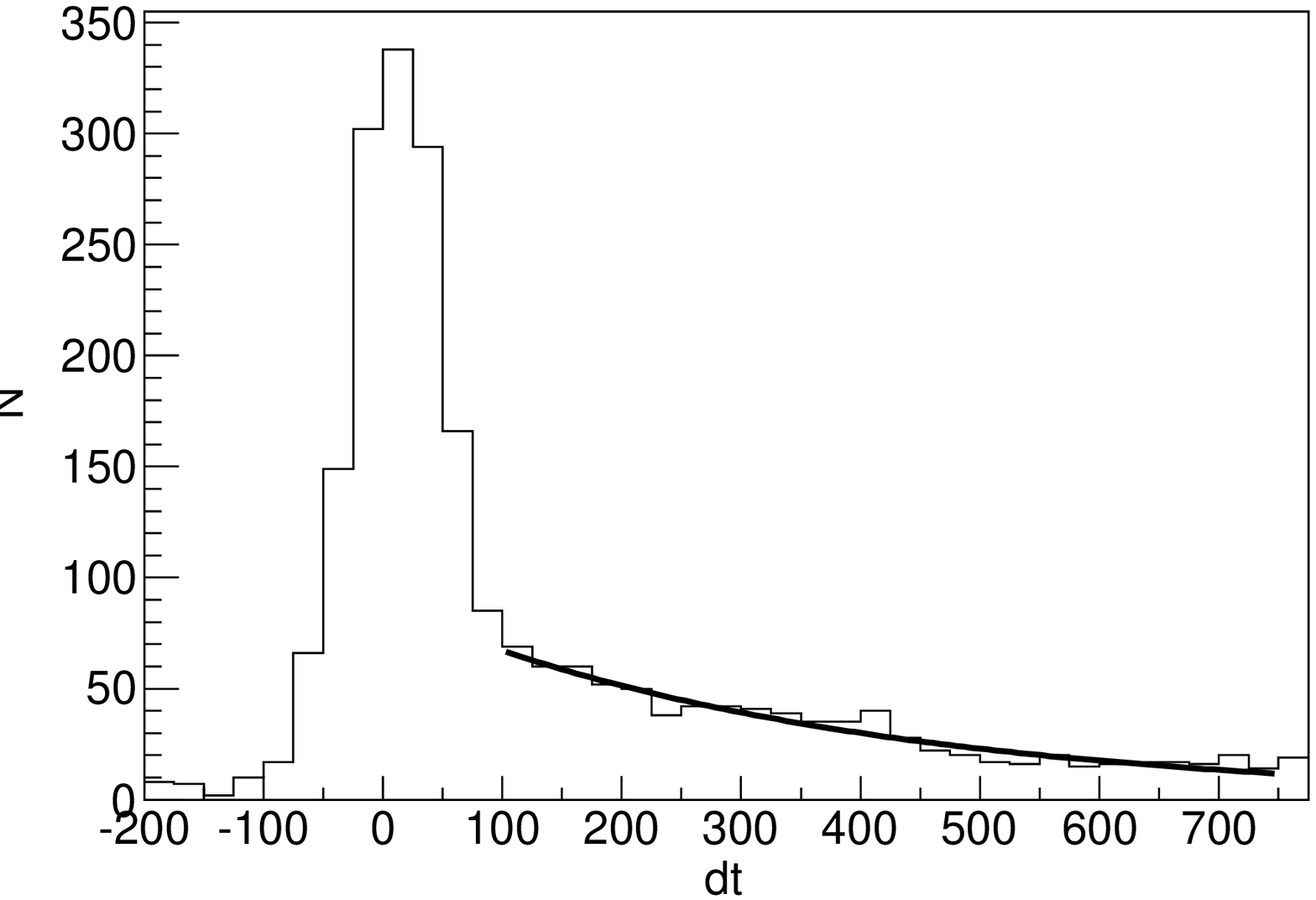}\\
	\vspace{-4.8cm}
	\hspace{5.7cm} (c) \hspace{.42\textwidth} (d)
	\\
	\vspace{4.8cm}
	\end{center}
	\caption{(Color online) Spectra for gamma ray energies in coincidence with identified protons.
		 (a): Doppler-corrected energy spectrum. (b): The  
		 laboratory gamma ray energy vs. the time difference 
		 between the detected proton and the detected gamma 
		 $(\Delta t = t_\gamma - t_\text{p})$. (c): The 
		 laboratory gamma ray energy spectrum before the 
		 reaction ($\Delta t < \SI{-75}{ns}$) (red (gray)) and after 
		 the reaction ($\Delta t > \SI{100}{ns}$) (black).
		 (d): The time difference between the detected gamma and
		 the detected proton for gamma rays in the peak at
		 $\SI{509.6}{keV}$ along with the fit to an exponential 
		 decay used to determine the lifetime of the $0^+_2$ state.
	}
	\label{fig:EgammaHE}
\end{figure*}

\subsection{Particles stopped in the $E$-detector}
\label{SS:HE}
Fig.~\ref{fig:dEE} shows a $\Delta E-E$ plot for one strip in one of
the detectors covering the forward laboratory angles. Protons, deuterons,
tritons and $\alpha$-particles are easily identified. A gate is made to
select the protons. Similar plots and gates have been made for each
strip separately.

The excitation energy spectrum for $^{12}$Be is obtained using the
kinematics of the identified protons, Fig.~\ref{fig:Eex}a.
A large part of the protons stems from reactions on C and H in the
target. The background is determined by analyzing the data from the 
runs on the pure carbon and the regular polyethylene target, and
is indicated by the green and yellow line in Fig.~\ref{fig:Eex}a. 
Fig.~\ref{fig:Eex}b shows the $^{12}$Be excitation energy spectrum
with the backgrounds subtracted. The ground
state of $^{12}$Be is clearly identified as the peak at $\SI{0}{MeV}$. There
is still a small additional background component at $\SI{1}{MeV}$, which might extend into the ground state peak. 
Only the background from C and H in the target are taken into account when
determining the ground state differential cross section, and the additional background will lead to an extra
uncertainty in the final spectroscopic factors, section~\ref{S:Results}. The $2^+_1$- and the $1^-_1$ states are also
visible in the spectra, but the energy resolution is too poor to separate the two or to see the $0^+_2$ state at $\SI{2.24}{MeV}$ and gates on gamma-ray energies are required
to identify these states. Gates on the gamma-ray energies have been determined from the spectra in Fig.~\ref{fig:EgammaHE}, which is described below. The gate on gamma-ray energies have been
applied and the spectra for the three excited states are shown in
Fig.~\ref{fig:Eex}c. Fig.~\ref{fig:Eex}d shows the total excitation energy
spectrum with efficiency corrected background and spectra with gates on gamma-ray energies applied
subtracted. This spectrum should thus represent the ground state of $^{12}$Be.
Only few events with $E^* >\SI{0.6}{MeV}$ are observed in Fig.~\ref{fig:Eex}d.
This shows that almost every event in the total spectrum can be described 
either from reactions on C and H in the target or from (d,p)-reactions.

The gates used for Fig.~\ref{fig:Eex}c are determined from two
spectra, Fig.~\ref{fig:EgammaHE}a+c. A gamma-ray energy spectrum is 
produced using gamma rays in coincidence with
the identified protons, Fig.~\ref{fig:EgammaHE}a. The energy is corrected
for Doppler-shift, due to the emission from a moving nucleus. Peaks
at $\SI{2103}{keV}$ and at $\SI{2722}{keV}$ are clearly seen. The two 
peaks are from the decay of the $2^+$- and the $1^-$ state to the ground
state respectively, see Fig.~\ref{fig:level} and Table~\ref{tab:Efficiency}.
Gates are set on the two peaks and excitation energy spectra of 
$^{12}$Be are generated using protons in coincidence with gamma rays 
within these gates, Fig.~\ref{fig:Eex}c (red and blue). The two spectra 
are scaled with $1/\epsilon$ from Table~\ref{tab:Efficiency} to take
the MINIBALL detection efficiency into account. The two peaks are situated 
at $2061\pm\SI{202}{keV}$ and $2658\pm\SI{192}{keV}$ respectively validating the
interpretation that the protons within the gate on the gamma-ray energies stem from the population
of the $2^+_1$ and the $1^-_1$ states.

The $0^+_2$ state is long lived, the lifetime of the state was
determined to be $\tau = 331\SI{(17)}{ns}$ by Shimoura et
al. \cite{Shimoura07}. The excited $^{12}$Be nuclei are either stopped
within the setup or far away from the MINIBALL detectors before
decaying. Only gamma rays from $^{12}$Be nuclei stopped within the setup can be
detected for the $0^+_2$ state. The nuclei are stopped in the forward 
silicon detectors or in the frame holding the detectors. This require an
outgoing angle larger than 7$^\text{o}$ for the $^{12}$Be nucleus, which
corresponds to center of mass angles between $71^\text{o}$ and $122^\text{o}$ for the protons.
The beam width gives a probability of the reaction happening off-center increasing
the required outgoing angle of $^{12}$Be for some events. This will lead to a drop in 
the detection efficiency for events with an outgoing $^{12}$Be angle close to $7^\text{o}$, leading 
to a larger uncertainty in the differential cross section around $70^\text{o}$ and $120^\text{o}$. 
The events populating the long lived
$0^+_2$ state can be identified by looking at time delayed
gamma rays. Fig.~\ref{fig:EgammaHE}b shows the time between the detected
gamma rays and the detected proton ($\Delta$t) against the laboratory
gamma energy. The laboratory gamma energy is used, 
as the gamma rays of interest come from stopped nuclei. Events with $\vert \Delta t \vert < \SI{75}{ns}$ is considered prompt decays and events with $\Delta t > \SI{100}{ns}$ is considered delayed gamma decays. The spectra in Fig.~\ref{fig:EgammaHE}c shows a projection onto the gamma energy axis before (red) 
and after (black) the reaction. A peak at $\SI{166}{keV}$ is present both before and after the reaction.
The peak also appears when carbon or regular polyethylene targets are used. The fact, that the peak is 
ever-present, even at times when no beam has hit the target¸ indicates that the peak stems from a background.
The origin of the peak is not known, but is expected to stem from long lived isotopes from a previous experiment. 
An excitation energy spectrum made with a gate on the 166 keV shows a flat distribution ranging from -4 MeV
to 3 MeV. This confirms the interpretation of a background peak. The reaction leads to
an increase in the overall background, but three new peaks emerge
after the reaction. Two narrow and one broad. The mean value and width
of the three peaks are determined using a Gaussian fit: $143.5\pm\SI{2.7}{keV}$,
$509.6\pm \SI{2.5}{keV}$ and $709\pm\SI{23}{keV}$. The first two are
identified as the decays of the
$0^+_2$ state, see Table~\ref{tab:Efficiency}. The last one stems from
decay of excited $^{72}$Ge within the MINIBALL detector. The germanium
isotopes are excited through inelastic scattering with neutrons and decay
subsequently \cite{Chas65,Jenkins09}. 

The two time delayed peaks have been used to determine the branching
ratio of the two decays:
\begin{eqnarray}
	BR_{0^+ \rightarrow 0^+} & = & 87.3\SI{(35)}{\%} \\
	BR_{0^+ \rightarrow 2^+} & = & 12.7\SI{(35)}{\%}
\end{eqnarray}
The large uncertainty on the detection efficiency of especially the
$\SI{143.5}{keV}$ gamma leads to a large uncertainty in the final result
of the branching ratio. The result is consistent with the values of
$82.3\SI{(15)}{\%}$ and $17.7\SI{(15)}{\%}$ determined earlier \cite{Shimoura07}.

The time signal also enabled a determination of the lifetime of the
$0^+_2$ state using the time difference
spectrum for the $\SI{511}{keV}$ gamma line, Fig.~\ref{fig:EgammaHE}d.
The spectrum is fitted to an exponential decay and gives a lifetime of:
\begin{equation*}
	\tau = 357\SI{(22)}{ns}.
\end{equation*}
The value is in fair agreement with the value ($\tau =
331(12)$ ns) determined by Shimoura et al. \cite{Shimoura07}.

The last peak (green) in Fig.~\ref{fig:Eex}c is made by
gating on the two time delayed gamma peaks. Again the
mean value of the excitation energy peak at $\SI{2190}{keV}$ validates
the two gamma peaks as stemming from the decay of the $0^+_2$ state. 
The spectra is scaled with a factor $1/0.63$ in addition to 
the $1/\epsilon$ from Table~\ref{tab:Efficiency}. This extra factor 
of $1/0.63$ stems from the additional timegate (d$t>\SI{100}{ns}$). For an 
exponential decay with a lifetime of $\SI{357}{ns}$, only $\SI{63}{\%}$ of the decays 
will be within the time window of [$\SI{100}{ns}$,$\SI{750}{ns}$].

To test for the presence of other components in the total excitation energy spectrum Fig.~\ref{fig:Eex}d shows the spectra in Fig.~\ref{fig:Eex}b with the spectra gated on the gamma-ray energies subtracted.
Most of the events above the ground state have disappeared, indicating that with the applied scaling, the spectra produced by gating on the gamma-rays can be used for cross section calculation without any major uncertainty in the overall amplitude. A small part of the total spectrum is still unaccounted for, which most likely stems from an unaccounted background, either from a small contamination of $^{22}$Ne in the beam or extra contaminations in the target. The possibility of some small extra components of the three excited states can not be ruled out though. Especially the $0^+_2$ state might not be fully described by the gate set on the gamma rays, due to the requirement of a stopped $^{12}$Be nucleus for the gamma detection. Furthermore, the 
detection efficiency ($\epsilon$) in Table~\ref{tab:Efficiency} is determined
for decays occurring at the target position not in the detectors at the end
of the setup, which could lead to a slight change in the scaling. 
Hence, the additional background leads to an extra uncertainty in the 
absolute amplitude of the cross sections for the two $0^+$ states. This 
is reflected in the uncertainties of the final spectroscopic factors, 
Table~\ref{tab:spec1}.

\subsection{Search for further bound states}
The extra data not accounted for by the excitation energy
 spectra gated on gamma-rays could also stem from a yet unseen bound state in $^{12}$Be. 
A bound $0^-_1$ state is predicted in a three-body model \cite{Rome208}. 
The excitation energy of the state is estimated to be between 
$\SI{2.1}{MeV}$ and $\SI{3.1}{MeV}$ in the model. Any $0^-_1$ state 
above the $1^-_1$ state can be ruled out by the data presented here. 
All events with excitation energy above $\SI{2.7}{MeV}$ are described 
by the spectra gated on gamma rays, Fig.~\ref{fig:Eex}d. Furthermore, there is no gamma 
line between $\SI{100}{keV}$ and $\SI{400}{keV}$ in Fig.~\ref{fig:EgammaHE}a.
The only peak present is the unknown background peak at $\SI{166}{keV}$. 
This narrows the energy search to the interval between $\SI{2.1}{MeV}$ 
and $\SI{2.7}{MeV}$. This interval can be further narrowed down. 
A $0^-_1$ state between the $2^+_1$ and the $1^-_1$ state would mainly 
decay to the $2^+_1$ state with an M2 transition. The state will then 
be long lived with a lifetime comparable to the $0^+_2$ state. 
Hence a peak between $\SI{200}{keV}$ and $\SI{500}{keV}$ should emerge 
in the black spectrum shown in Fig.~\ref{fig:EgammaHE}c, like the 
$\SI{511}{keV}$ and the $\SI{143.5}{keV}$ lines. No extra peak is seen 
in the spectra. From this we can limit the possible energy range 
for a bound $0^-_1$ state to:
\begin{equation}
	E^* \in [\SI{2.1}{MeV},\SI{2.2}{MeV}].
\end{equation}
From Fig.~\ref{fig:Eex}d we can determine the population strength for an 
additional state at $\SI{2.15}{MeV}$ to be more than a factor of 10 less
than the $1^-_1$ state. This would not be the case for a $0^-$ state, 
that only differs from the $1^-_1$ state in spin coupling. Therefore, 
it is very unlikely, that a bound $0^-$ state exists in $^{12}$Be.
\begin{figure*}[t]
	\begin{center}
	\psfrag{1+}[cc]{{\normalsize \hspace{1cm}$1^- \rightarrow 0^+_1$}}
	\psfrag{2+}[bc]{{\normalsize \hspace{1cm}$2^+ \rightarrow0^+_1$}}
	\psfrag{N}[bb]{{\normalsize counts/keV}}
	\psfrag{N2}[bb]{{\normalsize counts/keV}}
	\psfrag{E}[cc]{{\normalsize $E_\gamma$ [MeV]}}
	\psfrag{0+1}[bc]{{\normalsize \hspace{1cm}$0^+_2 \rightarrow 2^+$}}
	\psfrag{0+2}[tb]{{\normalsize \hspace{1cm}$0^+_2 \rightarrow0^+_1$}}
	\psfrag{Ge}[bc]{{\normalsize $^{72}$Ge$^*$}}
	\includegraphics[width=0.45\textwidth]{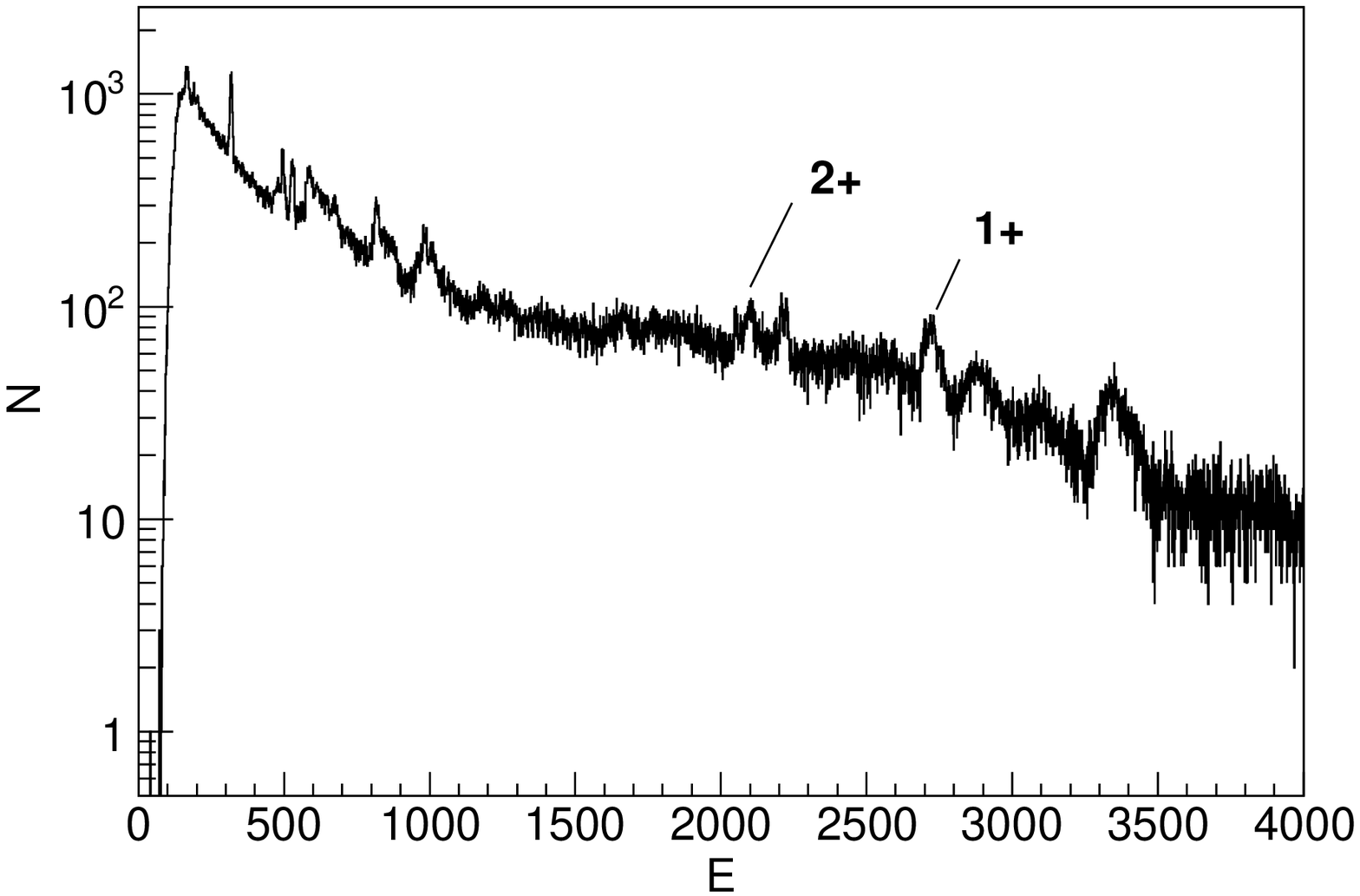}
	\includegraphics[width=0.45\textwidth]{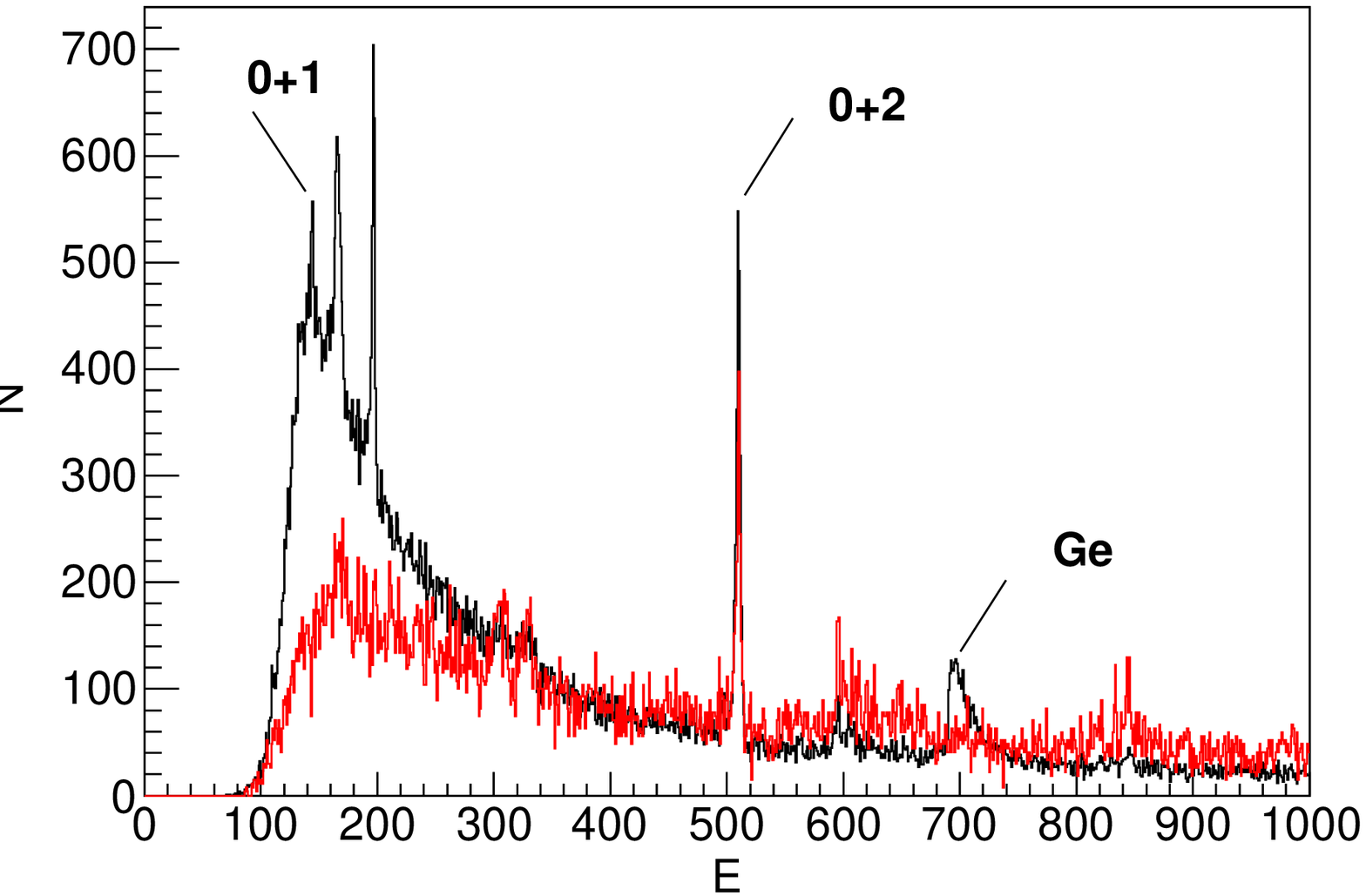}\\
	\vspace{-4.8cm}
	\hspace{5.7cm} (a) \hspace{.42\textwidth} (b)
	\\
	\vspace{4.8cm}
	\end{center}
	\caption{(Color online) $E_\gamma$-spectra for gamma rays in coincidence with
          unidentified charged particles stopped in the $\Delta
          E$-detectors. (a): Doppler-corrected energy spectrum.
          (b): Laboratory gamma energy spectra for gamma rays
	  emitted before (red (gray)) and after (black) the reaction.
	  The time gates are similar to the ones in Fig.~\ref{fig:EgammaHE}c.}
	\label{fig:LEgamma}
\end{figure*}
\begin{figure}[h]
	\psfrag{N}[bb]{{\normalsize counts/$\SI{50}{keV}$}}
	\psfrag{E}[cc]{{\normalsize $E^*$ [MeV]}}
	\includegraphics[width=0.5\textwidth]{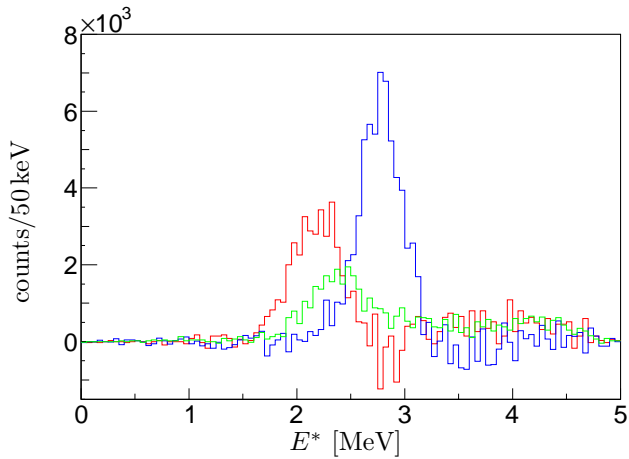}
	\caption{(Color online) The excitation energy of $^{12}$Be determined from the
		momentum of particles stopped in the $\Delta E$-detectors
		in the forward angles and gated on the gamma lines shown in
		Fig.~\ref{fig:LEgamma}. red (dark gray): $2^+_1$ 
		 ($E_\text{peak}=\SI{2061}{keV}$), green (light gray): $0^+_2$ 
		 ($E_\text{peak}=\SI{2190}{keV}$) and blue (black): $1^-_1$
		 ($E_\text{peak}=\SI{2658}{keV}$).}
	\label{fig:LEEex}
\end{figure}

\subsection{Particles stopped in the forward $\Delta E$-detector}
\label{SS:LEfor}

Studying the particles stopped in the $\Delta
E$-detectors is the next step. We first consider the forward laboratory angles. All protons
producing $^{12}$Be in the ground state go through the $\Delta
E$-detector in forward angles, see Fig.~\ref{fig:TREX}b. This leaves
only the protons to excited states in $^{12}$Be to be identified. The
population of the excited states can be determined using the same
gates as in the previous section. Fig.~\ref{fig:LEgamma}a shows the 
Doppler-corrected gamma rays in coincidence with particles stopped in the
 $\Delta E$ and Fig.~\ref{fig:LEgamma}b shows the laboratory 
gamma energy before (red) and after (black) the reaction. The plots
are similar to Fig.~\ref{fig:EgammaHE}a+c. More peaks appears in the
Doppler-corrected spectrum. These gamma rays stem mainly from inelastically scattered
$^{11}$Be ($E_\gamma = \SI{320}{keV}$), excited states in
$^{10}$Be populated in (d,t)-reactions ($E_\gamma = \SI{2590}{keV}$,
$\SI{2812}{keV}$ and $\SI{3367}{keV}$) and reactions on C and H. The 
two peaks at $\SI{2096}{keV}$ and $\SI{2723}{keV}$ are still present 
and easily separable from other gamma lines. 

Comparing the laboratory gamma energy
spectra before and after the reaction time shows the same appearance
of the three peaks mentioned in the previous section. Two things
should be noted. A fourth peak at $\SI{197}{keV}$ appears after
the reaction point. The peak stems from decays in $^{19}$F populated
in reactions of $^{11}$Be on $^{12}$C in the target. Secondly, a
significant peak at $\SI{511}{keV}$ is seen before the reaction
time. This indicates a non-negligible background from positrons
within the ISOLDE experimental hall. This leads
to a significant background when using the gate, which is taken into
account when producing the excitation energy spectrum gating on gamma rays,
Fig.~\ref{fig:LEEex}. All three spectra produced with a gate on gamma-ray energies in
Fig.~\ref{fig:LEEex} are peaked at the correct excitation energies and
the background which remains is negligible.
\begin{figure*}[t]
	\psfrag{N}[bb]{{\normalsize counts/$\SI{50}{keV}$}}
	\psfrag{E}[cc]{{\normalsize $E^*$ [MeV]}}
	\includegraphics[width=0.45\textwidth]{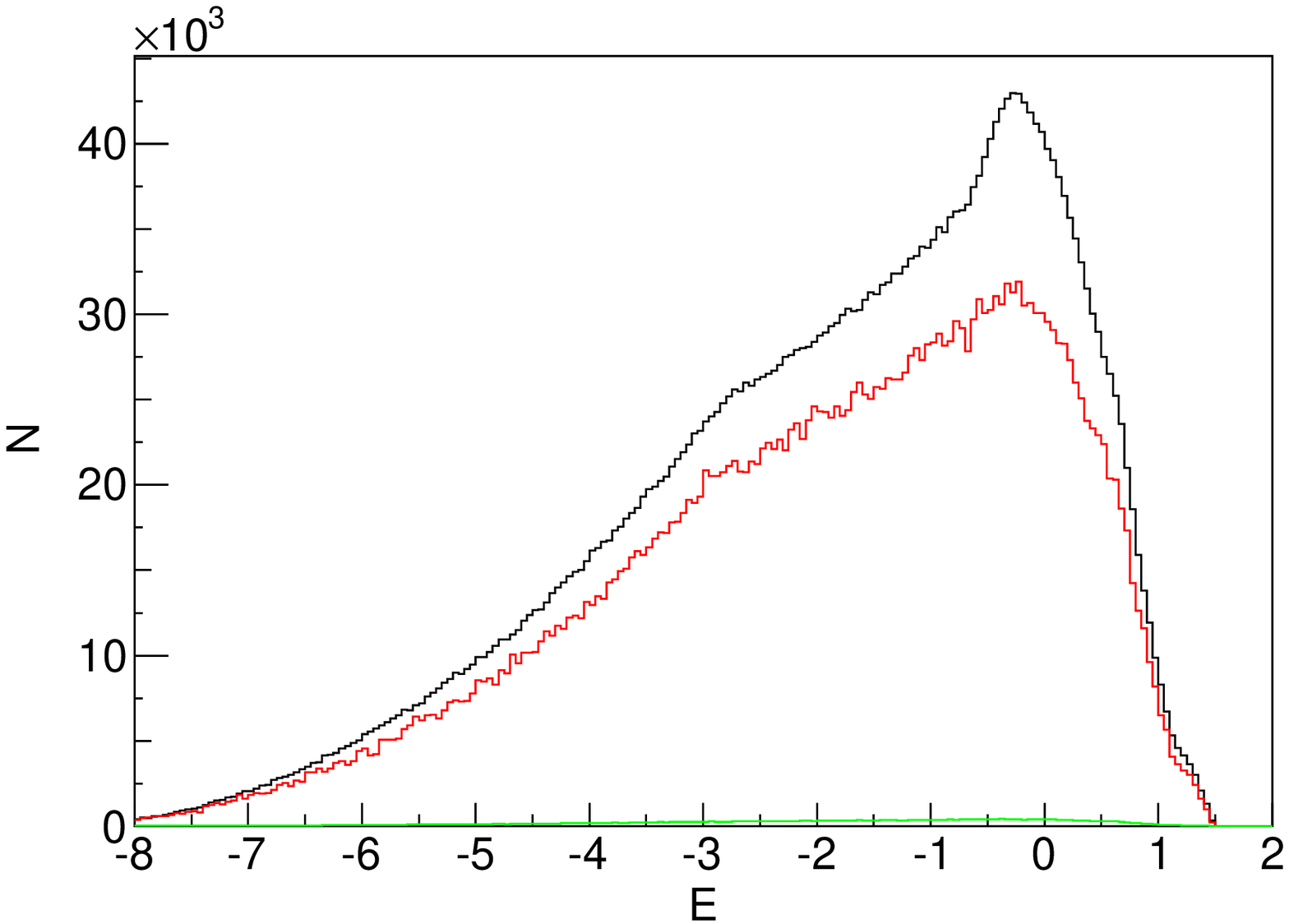}
	\includegraphics[width=0.45\textwidth]{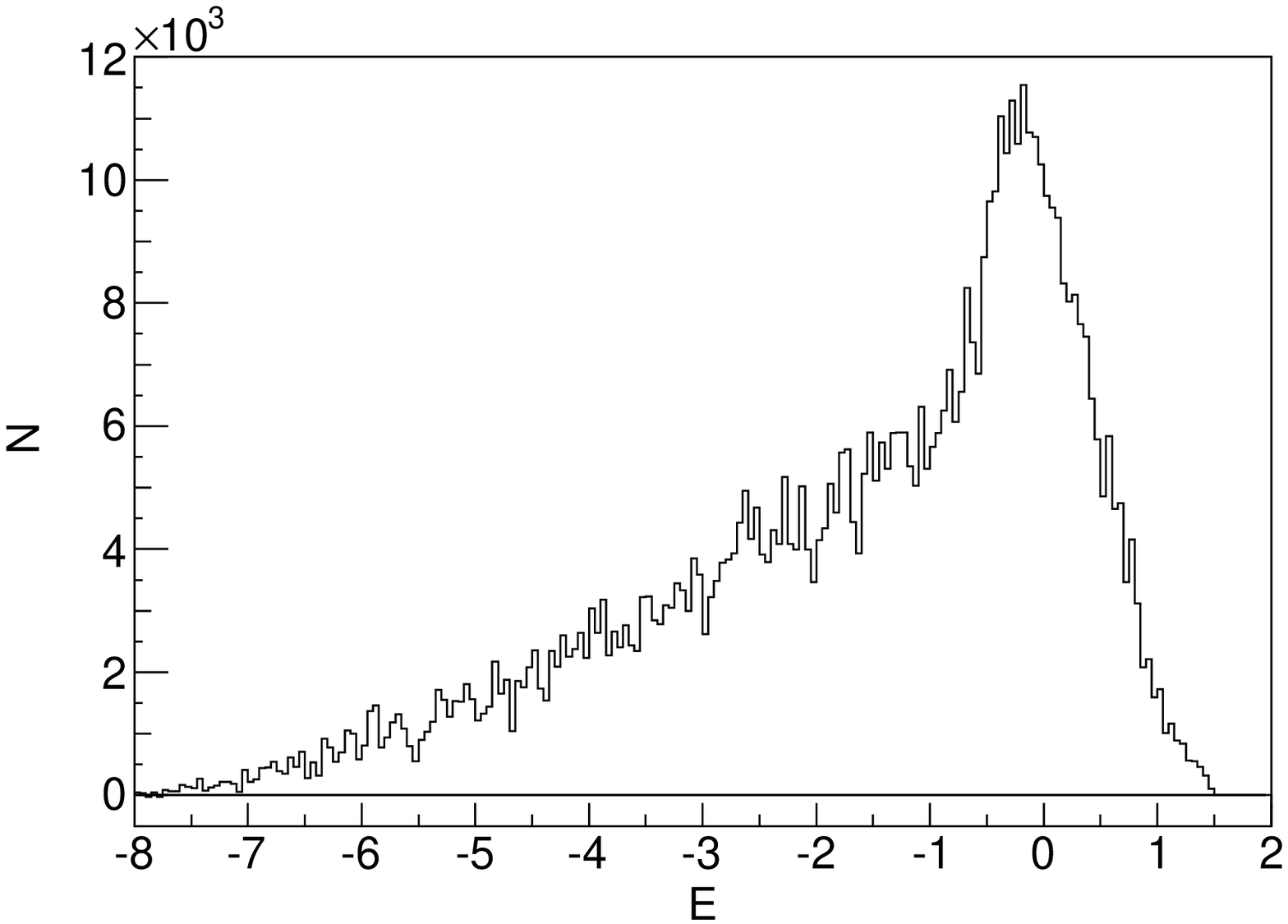}\\
	\vspace{-4.8cm}
	\hspace{5.7cm} (a) \hspace{.42\textwidth} (b)
	\\
	\vspace{4.8cm}
	\caption{(Color online) Excitation energy spectra of $^{12}$Be calculated 
	  for particles in backward laboratory angles. (a): The total
	  excitation energy spectrum (black) along with the background
	  spectra from runs on the carbon target (red (dark gray)) and the CH$_2$
	  target (green (light gray)). (b): The excitation energy spectrum with the two
   	  background spectra subtracted.}
	\label{fig:BackEex}
\end{figure*}
\begin{table*}[t]
\begin{tabular}{|l|l|cccccccccccc|}
\hline
Channel & set & $V_0$ & $r_0$ & $a_0$ & $\delta_1$ & $\delta_2$ & $W_v$ & $W_d$ & $r_I$ & $a_I$ & $V_{so}$ & $r_{so}$ & $a_{so}$ \\
\hline
$^{12}$Be+p & I+II+III & 57.8 & 1.25 & 0.25 & 0 & 0 & 0 & 8.08 & 1.4 & 0.22 & 6.5 & 1.25 & 0.25 \\
 & IV & 58.59 & 1.12 & 0.67 & 0 & 0 & 0.85 & 5.26 & 1.3 & 0.51 & 5.53 & 0.90 & 0.59 \\
 \hline
 & I & 124.7 & 0.9 & 0.9 & 0 & 0 & 0 & 4.38 & 2.452 & 0.264 & 6.0 & 0.9 & 0.9 \\
$^{11}$Be+d  & II &  120.18 & 0.9 & 0.9 & 0.84 & 1.27 & 0 & 19.535 & 2.452 & 0.264 & 6.0 & 0.9 & 0.9 \\
 & III & 118.0 & 0.87 & 0.91 & 0 & 0 & 0 & 5.80 & 1.57 & 0.78 & 5.80 & 0.87 & 0.91 \\
 & IV & 80.53 & 1.17 & 0.8 & 0 & 0 & 5.19 & 4.71 & 1.56 & 0.8 & 3.54 & 1.23 & 0.81\\
\hline
$^{1}$H+n & & 72.15 & 1.484 &  & & & & & & & & & \\
\hline
$^{11}$Be+n & & 54.14 & 1.35 & 0.9 & & & & & & & 8.50 & 1.35 & 0.9\\
\hline
\end{tabular}
\caption{Parameters for the six optical potentials used in the four (I-IV) DWBA calculations along with the two binding potentials for $^2$H and $^{12}$Be. All potentials have a Wood-Saxon shape except the $^{1}$H+n potential, which is a Gaussian shape.}
\label{tab:Potentials}
\end{table*}

\subsection{Particles stopped in the backward $\Delta E$-detector}
\label{SS:LEback}
The last part of the analysis concerns the backward laboratory
angles. Protons producing $^{12}$Be in the ground state are the only
ones detected in the backward angles, due to the high lower energy
detection threshold, Fig.~\ref{fig:TREX}b. An excitation energy spectrum is
made from all particles stopped in the $\Delta E$ detector in backward
angles. All particles are assumed to be protons, Fig.~\ref{fig:BackEex}a. 
Background from C and H in the target are subtracted using 
the runs on pure carbon and regular polyethylene targets, Fig.~\ref{fig:BackEex}b. The ground state is clearly seen, but a significant background is
still present. The background leads to a larger uncertainty
in the determined differential cross sections for the small center of
mass angles.

\section{Results}
\label{S:Results}
\begin{figure}[h]
	\center
	\psfrag{x}[tb]{$\frac{\text{d}\sigma}{\text{d}\Omega}$ [mb/sr]}
	\psfrag{t}[bb]{$\theta_\text{cm}$ [deg]}
	\includegraphics[width=.5\textwidth]{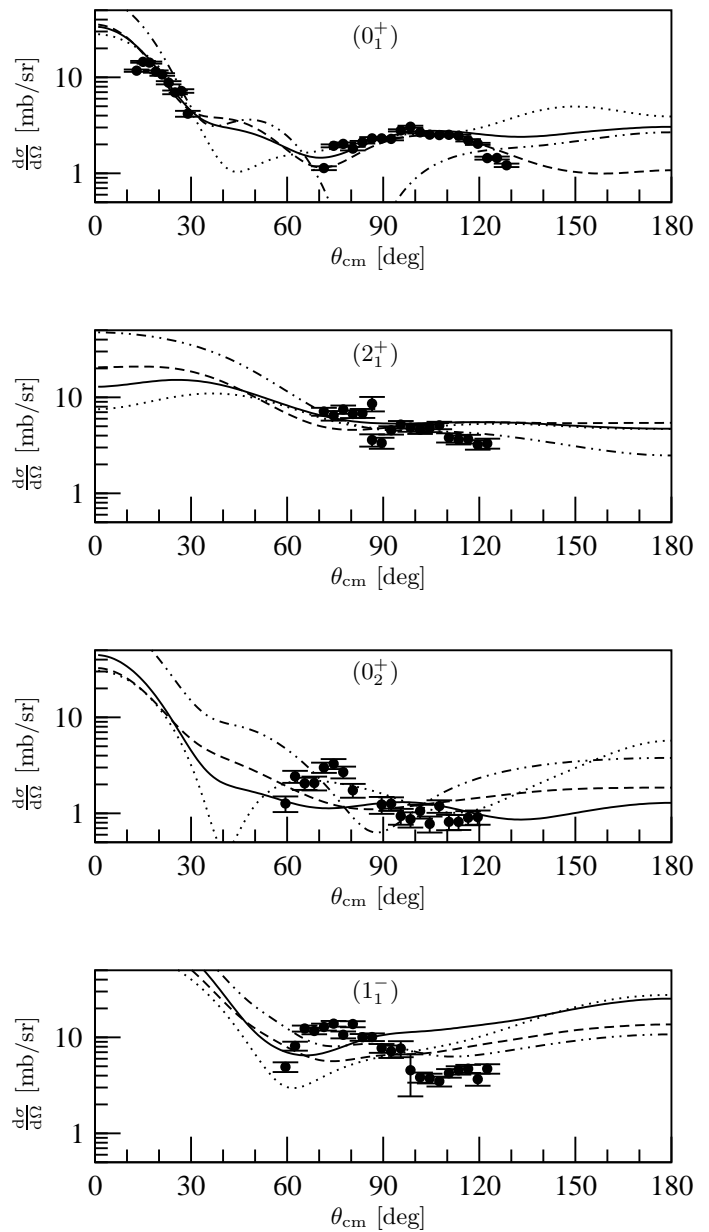}
	\put(-130,450.5){$(0^+_1)$}
	\put(-130,330){$(2^+_1)$}
	\put(-130,209.5){$(0^+_2)$}
	\put(-130,89){$(1^-_1)$}
	\caption{The experimentally determined differential cross
          sections for the four bound states in $^{12}$Be (dots). 
	  The DWBA calculations are plotted on top of the
          experimental data; set I: full line (---), II: dashed (- -
          -), III: dotted (...) and IV: dash-dotted (.-.).}
	\label{fig:xsec}
\end{figure}

The experimental differential cross sections for the (d,p)-population
of the four bound states are determined by comparing the excitation energy 
spectrum, determined in section~\ref{S:Analysis}, with a geant4
\cite{geant4} simulation. This simulation was done using the
g4miniball package \cite{Bildstein12}.  The excitation energy
spectra produced using gates on the gamma-ray energies from Fig.~\ref{fig:Eex}c and \ref{fig:LEEex} are used
for the excited states ($2^+_1$, $0^+_2$ and $1^-_1$). The total
excitation energy spectra from Fig.~\ref{fig:Eex} and \ref{fig:BackEex} are
used for the ground states. The experimental differential cross
sections are shown in Fig.~\ref{fig:xsec} (dots).

Theoretical calculations of the differential cross sections are needed
in order to extract conclusions from the experimental
results. This step is complex in our case, partly because the
``forward peak'' is only covered for the ground state transition,
partly because both initial nuclei --- the deuteron and $^{11}$Be ---
are loosely bound systems. The theory for (d,p) reactions is still
being refined for challenging cases like this and is often making use
of continuum discretized coupled channel (CDCC) calculations, see
\cite{Kee04,Moro09,Mukh11} and references therein. The concept of a
spectroscopic factor, often used earlier as the key quantity to be
extracted from experiment, has also been questioned during the last
decade, see Mukhamedzhanov \cite{Mukh11} and Jennings \cite{Jen11} for
recent overviews, asymptotic normalization coefficients (ANC)
\cite{Mukh99,Timo03,Mukh11,Teng11} have been used instead. A proper
theoretical analysis of our data is beyond the scope of the present
paper so we shall here just briefly indicate what can be learned from
a standard approach to allow comparison with earlier experimental work
\cite{Kanungo10}.

The differential cross sections are compared to distorted wave Born
approximation (DWBA) calculations performed with FRESCO
\cite{Thompson88,Thomp09}. Four DWBA calculations are performed using
the parameters from Table~\ref{tab:Potentials}. The interaction
potentials are of the form:
\begin{eqnarray*}
	\label{eq:UOM}
	V(r) & = & -V_0f(x_0) - i\left(W_v f(x_I) - W_d \frac{\text{d}f((x_I))}{\text{d}x_I} \right) \\
	&+ &V_{so}\frac{\hbar^2}{m_\pi c} \frac{1}{r} \frac{\text{d}f((x_{so}))}{\text{d}x_{so}} (\vec{L}\cdot \vec{s})
\end{eqnarray*}
where $f(x)$ is the Wood-Saxon:
\begin{eqnarray*}
	\label{eq:fWS}
	f(x_i) & = & \frac{1}{1+\exp(x_i)} \\
	\label{eq:xOM}
	x_i  & = & \frac{r- r_i A^{1/3}}{a_i}.
\end{eqnarray*}

The $^{12}$Be+p potentials are taken from reference \cite{PotI} (set I+II+III)
and \cite{PotIV} (set IV). The first $^{11}$Be+d potential is
calculated from generalized parameters given by Satchler et
al.\ \cite{Satchler66} (set I). The depths of this potential is
modified for the second set (set II) and a deformation taken from
Hussein et al.\ \cite{Hussein08} is added. The second potential fits
better the elastic scattered deuterons (not investigated
here). The last two $^{11}$Be potentials are taken from Fitz et al.\ \cite{Fit67} and Han
et al.\ \cite{Han06}. These two potentials were used in combination with
the two $^{12}$Be+p potentials to investigate the differential cross
sections from the $^{11}$Be(d,p)$^{12}$Be experiment performed at TRIUMF by Kanungo et al. \cite{Kanungo10}.

The two binding potentials are taken from Austern et al. \cite{Austern87} for 
$^{1}$H+n and Nunes et al. \cite{Nunes96} for $^{11}$Be+n. The $^{11}$Be+n potential has a Wood-Saxon shape and the $^{1}$H+n has a Gaussian form:
\begin{eqnarray*}
	V(r)  & = & V_o\exp\left[\left(r/r_0\right)^2\right].
\end{eqnarray*}
The parameters can be seen in Table~\ref{tab:Potentials}.

The theoretical differential cross sections for each of the four sets
of parameters are plotted in Fig.~\ref{fig:xsec}. The deduced
spectroscopic factors from the four calculations are given in
Table~\ref{tab:spec1}. The spectroscopic factors are given along 
with spectroscopic factors determined by Kanungo et al. \cite{Kanungo10}
and from two theoretical models: a shell model calculation made 
by H. Fortune et al.\ \cite{Fortune11} and a three-body model 
calculation made by E. Garrido et al.\ \cite{Garrido13}.  Effects of 
core excitation are not yet included in the three-body model \cite{Garrido13}.

The cross sections from set IV can only reproduce the shape of the
$2^+_1$ cross section and the validity of these parameters is very
questionable. The first three sets reproduces the ground state well
and to a large extent the $2^+_1$ state. Only set III can reproduce
the shape for the $0^+_2$ state and none of the potentials can
reproduce the shape for the $1^-_1$. The low binding energy and the
possible two-halo structure of the $1^-_1$ state is expected to play
an important role in the reaction mechanism to the $1^-_1$. The
overall agreement is not satisfactory, but as mentioned above this is
not too surprising and a better description of the reaction mechanisms
is needed where effects due to the halo structure of the two initial
nuclei, e.g.\ break-up of the halo \cite{Moro09,Moro12}, are included.
\begin{table*}[t]
\center
\begin{tabular}{|l|c|c|c|c|c|c|c|}
\hline
State & set I & set II & set III & set IV & 
Ref.\ \cite{Kanungo10} & Ref.\ \cite{Fortune11} & Ref.\ \cite{Garrido13}\\
\hline
$0^+_1$ & 0.15$\substack{+0.03 \\ -0.05}$ & 0.25$\substack{+0.05 \\ -0.08}$ & 0.15$\substack{+0.03 \\ -0.05}$ & [0.30$\substack{+0.20 \\ -0.22}$] &  0.28$\substack{+0.03 \\-0.07}$ & 0.78 & 0.60\\
$2^+_1$ & 0.15(5) & 0.30(10) & 0.075(25) & 0.40(10) & 0.1$\substack{+0.09 \\-0.07}$ & 0.52 & (0.35)\\
$0^+_2$ & [0.40$\substack{+0.14 \\ -0.10}$] & [0.32$\substack{+0.12 \\ -0.09}$] & 0.40$\substack{+0.13 \\ -0.09}$ & [0.95$\substack{+0.43 \\ -0.36}$] & 0.73$\substack{+0.27 \\-0.40}$  & 0.37 & 0.07\\
$1^-_1$ & [0.55(20)] & [0.50(20)] & [0.27(15)] & [0.85(35)] & $\approx 0.35$ & -- & 0.50 \\
\hline
\end{tabular}
\caption{Spectroscopic factors for the four bound states in
  $^{12}$Be. The spectroscopic factors are given for each set of
 parameters shown in Table~\ref{tab:Potentials} along with 
 spectroscopic factors from a $^{11}$Be(d,p) experiment performed 
 at TRIUMF \cite{Kanungo10}, a shell model calculation \cite{Fortune11}
 and a three-body calculation \cite{Garrido13}. Square brackets
 indicate cases where the angular shapes do not match.}
\label{tab:spec1}
\end{table*}

The deduced spectroscopic factors are highly model dependent as seen
in table III. Especially set IV gives values, which are not consistent 
with any of the other sets. The validity of the values from set IV has 
already been questioned, due to the discrepancy in the angular shapes. 
The strong disagreement between set III and IV is in contradiction with 
the result found in [25], where the two sets is claimed to provide 
consistent results. The spectroscopic factors for the excited states 
found by set III are consistent with the values found in [25]. For the 
excited $0^+_2$ state the previous determination carried a large uncertainty 
since the $2^+_1$ and the $0^+_2$ states could not be separated in [25]. 
The states are identified and separately analyzed in this experiment and 
this should provide a more reliable value for the $0^+_2$ state. The factor 
of two between the two ground state values is not understood. 
 
The experimental spectroscopic
factor is, in contrast to the theoretical ones, larger in the
excited $0^+$ state than the ground state; this and the overall large
disagreement between the experimental determined spectroscopic factors
and the theoretical ones is still to be understood. 

\section{Summary and conclusion}
\label{S:Conclusion}
The combined power of the T-REX and MINIBALL arrays allows to identify
individual final states in the $^{11}$Be(d,p)$^{12}$Be reaction. All
previously known excited bound states have been seen through
observation of gamma rays from their decay. For the decay of the
$0^+_2$ state the lifetime has been measured to $\tau = 357\SI{(22)}{ns}$
and the branching ratios of the decays to the ground state and the
$2^+_1$ state have been determined to $BR_{0^+ \rightarrow 0^+} =
87.3\SI{(35)}{\%}$ and $BR_{0^+ \rightarrow 2^+} = 12.7\SI{(35)}{\%}$
respectively. These values are in good agreement with previously
determined values. No indications for new bound states were seen,
there are in particular no indications for the presence of a bound
$0^-_1$ state. The excitation energy of such a bound state has been limited to be between the $2^+_1$ and the $0^+_2$ state in an interval of only $\SI{100}{keV}$. The amount of unaccounted data within this interval will lead to a population strength much below that expected for a bound $0^-_1$ state. Hence a fifth bound state in $^{12}$Be can be ruled out.

Differential cross sections have been extracted over a large angular
range ($60{^\text{o}}$ to $120{^\text{o}}$ in the center-of-mass system) and
compared to four different sets of DWBA calculations in order to
determine spectroscopic factors. None of the DWBA calculations could
reproduce all of the experimental differential cross sections. The
difference between the experimental and theoretical differential cross
sections is large, especially for the high lying levels.  This may be
due to the loosely bound neutrons in the initial states and the
suggested halo structure of some of the final states, factors which
are known to affect the reaction mechanism. More refined calculations
must be made for the theoretical differential cross sections, e.g.\
CRC found to be essential to describe the $^8$Li+$^2$H reaction
at a similar energy \cite{Teng11}. The current disagreement between
theoretical and experimental spectroscopic factors may be due to the
simplicity of the DWBA calculations, but it is disturbing that the
relative strength of transitions to the two $0^+$ states has opposite
trends for theory and experiment. It will clearly be important to go
beyond the simple theoretical treatment presented here.

\texttt{Acknowledgments:}
This work was supported by the European Union Seventh Framework
through ENSAR (contract no. 262010), by the BMBF under the contracts
06MT7178, 06MT9156, 05P09PKCI5, 05P12PKFNE, 06DA9036I and 05P12RDCIA,
by the Spanish MICINN under the contracts FPA2010-17142 and FPA2012-34332,
by the FWO-Vlaanderen (Belgium), by GOA/10/010 (BOF KU Leuven), by
the Interuniversity Attraction Poles Programme initiated by the 
Belgian Science Policy Office (BriX network P7/12), by the United 
Kingdom Science and Technology Facilities Council, by a Marie Curie 
Intra-European Fellowship of the European Community's 7$^\text{th}$ Framework 
Programme under contract number PIEF-GA-2008-219175, and by 
Maier-Leibnitz-Laboratorium, Garching. The authors would like to thank
Aksel S. Jensen and Eduardo Garrido for valuable inputs and discussions
regarding the nuclear structures. We would also like to thank Antonio M.
Moro for his input regarding the theoretical calculations of the differential
cross sections.


\end{document}